\documentclass[preprint,prd,nofootinbib,showpacs]{revtex4}
\usepackage{graphicx}
\usepackage{epsfig}
\usepackage{bm}
{
{\newcommand{\gsim}{\mbox{\raisebox{-.6ex}{~$\stackrel{>}{\sim}$~}}}

\begin{document}

\title{Perturbative Bottom-up Approach for Neutrino Mass Matrix in Light of Large $\theta_{13}$ and Role of Lightest Neutrino Mass}

\author{Rupak Dutta${}$}
\email{rupak@iith.ac.in}
\author{Upender Ch${}$}
\email{ph11p1003@iith.ac.in}
\author{Anjan K. Giri${}$}
\email{giria@iith.ac.in}
\author{Narendra Sahu${}$}
\email{nsahu@iith.ac.in}
 
\affiliation{
Indian Institute of Technology Hyderabad, Yeddumailaram 502205, AP, India.
}
\preprint{}

\begin{abstract}
We discuss the role of lightest neutrino mass ($m_0$) in the neutrino mass matrix, defined in a flavor basis, through 
a {\it bottom-up} approach using the current neutrino oscillation data. We find that if $m_0 < 10^{-3} {\rm eV}$, then the 
deviation $\delta M_\nu$ in the neutrino mass matrix from a tree-level, say tribimaximal neutrino mass matrix, does not 
depend on $m_0$. As a result $\delta M_\nu$'s are exactly predicted in terms of the experimentally determined quantities 
such as solar and atmospheric mass squared differences and the mixing angles. On the other hand for $m_0 \gsim 10^{-3} {\rm eV}$, 
$\delta M_\nu$ strongly depends on $m_0$ and hence can not be determined within the knowledge of oscillation parameters alone. 
In this limit, we provide an exponential parameterization for $\delta M_\nu$ for all values of $m_0$ such that it can 
factorize the $m_0$ dependency of $\delta M_\nu$ from rest of the oscillation parameters. This helps us in finding 
$\delta M_\nu$ as a function of the solar and atmospheric mass squared differences and the mixing angles for all values of $m_0$. We use 
this information to build up a model of neutrino masses and mixings in a top-down scenario which can predict large $\theta_{13}$ perturbatively. 

\end{abstract}

\maketitle

\section{Introduction}
Compelling evidences of neutrino oscillations observed in solar, atmospheric and reactor experiments indicate that 
neutrinos are massive and hence they mix with each other~\cite{pontecorvo}. In the basis where the charged lepton 
masses are real and diagonal, the mixing matrix is given by~\cite{pmns}:
\begin{eqnarray}
\label{upmns}
U_{{\rm PMNS}} = \left( \begin{array}{ccc}
c_{12}\,c_{13} & s_{12}\,c_{13} & s_{13}\,e^{-i\delta_{\rm CP}} \\
-c_{23}\,s_{12} - s_{23}\,c_{12}\,s_{13}\,e^{i\delta_{\rm CP}} & c_{23}\,c_{12} - s_{23}\,s_{12}\,s_{13}\,e^{i\delta_{\rm CP}} & s_{23}\,c_{13} \\
s_{23}\,s_{12} - c_{23}\,c_{12}\,s_{13}\,e^{i\delta_{\rm CP}} & -s_{23}\,c_{12} - c_{23}\,s_{12}\,s_{13}\,e^{i\delta_{\rm CP}} & c_{23}\,c_{13} \end{array} \right)P,
\end{eqnarray}
where $c_{ij} = \cos\theta_{ij}$, $s_{ij} = \sin\theta_{ij}$ and $\delta_{\rm CP}$ is the Dirac CP violating phase on which 
the oscillation probability depends. The diagonal matrix $P = {\rm diag}(e^{i\phi_1},\,e^{i\phi_2},\,1)$, consists of two 
Majorana phases $\phi_1$ and $\phi_2$ which are not relevant in neutrino oscillation experiments~\cite{Bilenky,Langacker}. 
However, they affect lepton number violating amplitudes such as neutrinoless double beta decay. Thus in a flavor basis, where 
charged leptons are real and diagonal, the neutrino mass matrix is given by:
\begin{eqnarray}
M_{\nu} = U^{\ast}_{{\rm PMNS}}\,{\rm diag}(m_1,\,m_2,\,m_3)\,U^{\dagger}_{{\rm PMNS}},
\end{eqnarray}
where $m_1,\,m_2,\,m_3$ are the mass eigenvalues. 

In the last decade, data from solar and reactor neutrino experiments have provided information on the sign and
magnitude of $\Delta m_{\odot}^2$ and a precise value of $\theta_{12}$~\cite{sno,kamland}. The atmospheric parameters $|\Delta
m_{\rm atm}^2|$ and $\theta_{23}$ have been measured and their precision will be increased by T2K~\cite{t2k} and NO$\nu$A~\cite{nova}.
The sign of solar mass splitting $\Delta m_{\odot}^2$  is precisely known, while the sign of atmospheric mass splitting 
$\Delta m_{\rm atm}^2$ is still unknown. This opens up a possibility of whether neutrino masses follow normal ordering, 
i.e., $m_1 < m_2 < m_3$ or inverted ordering, i.e., $m_3 < m_1 < m_2$. In other words, the lightest mass, either $m_1$ (normal ordering) or 
$m_3$ (inverted ordering) is yet to be determined. Recent measurement from T2K~\cite{t2k2}, MINOS~\cite{minos}, Double 
Chooz~\cite{DC}, Daya Bay~\cite{daya}, and RENO~\cite{reno} confirms a non-zero value of $\theta_{13}$ at $5\sigma$ 
confidence level. This opens up a range of possibilities to measure the sign of the atmospheric mass splitting and the unknown CP 
phase $\delta_{CP}$. 

The absolute mass scale of neutrino is hitherto not known and can only be measured in a tritium beta decay experiment.
The KATRIN experiment, which will investigate the kinematics of tritium beta decay, aims to measure the neutrino mass with a 
sensitivity of $0.2\,{\rm eV}$~\cite{katrin}. At present the best upper limit at $95\%$ confidence level on the sum of the neutrino 
masses comes from the cosmic microwave background data and is given by~\cite{wmap9}: 
\begin{equation}
\sum_i m_i <  0.44 eV\,.
\end{equation} 
Once the absolute mass scale of the lightest neutrino, either $m_1$ in normal ordering or $m_3$ in inverted ordering, is 
determined one can reconstruct the neutrino mass matrix in the flavor basis using the experimental values of the elements 
of $U_{{\rm PMNS}}$ matrix. This may unravel exact flavor structure in the neutrino mass matrix.

In light of recent non-zero $\theta_{13}$, a large number of flavor models have been proposed by using the top-down 
approach, where one assumes a specific symmetry~\cite{symmetry_1,symmetry_2,symmetry_3,symmetry_4,symmetry_5,reviews} to explain the 
observed masses and mixings of the light neutrinos. However, it is worth exploring the symmetries of neutrino mass matrix 
through a data driven approach~\cite{data_drive_approach}. But as we discussed above the mass of lightest neutrino is yet 
to be known and therefore, the low energy neutrino data may not unravel the full flavour structure. 

In this paper we develop a perturbative bottom-up approach to unravel the flavor structure of neutrino mass 
matrix and discuss the role of lightest neutrino mass. We set the full neutrino mass matrix: $M_\nu=(M_\nu)_0 + \delta M_\nu$, 
where $\delta M_\nu$ is the perturbation around the tree-level mass matrix $(M_\nu)_0$, which is determined using some 
of the well known mixing scenarios such as tribimaximal~(TBM) mixing~\cite{tbm}, bimaximal~(BM) mixing~\cite{bm} and/or 
democratic~(DC) mixing~\cite{dcm}. Among all these, the TBM is closer to the experimentally observed mixing pattern and 
hence mostly studied. All these mixing scenarios, however, predict $\theta_{13}$ to be zero and hence ruled out by the 
recent measurement on the reactor neutrino angle. However, a perturbative approach to realize a large value of $\theta_{13}$ 
is still a viable option. So, we assume that the non-zero value of $\theta_{13}$ 
is generated perturbatively. Using the $3\sigma$ range of values of the elements of $U_{{\rm PMNS}}$ matrix we determine 
$\delta M_\nu$ as a function of the lightest neutrino mass, say $m_0$, where $m_0=m_1$ in the normal ordering and $m_0=m_3$ 
in the inverted ordering. In this way we find that for $m_0 < 10^{-3}$ eV, all the $\delta M_\nu$ are independent of $m_0$ and 
hence the lightest neutrino mass does not play any role in the perturbative determination of neutrino mass matrix. Such models, for 
example, can be generated in two right-handed neutrino extensions of the standard model (SM). However, for $m_0 \gsim 10^{-3}$ eV 
the lightest neutrino mass plays an important role in the perturbative determination of neutrino mass matrix. We factorize the $m_0$ 
dependency of $\delta M_\nu$ using an exponential parameterization: $\delta M_\nu (m_0) \propto \delta M_\nu|_{m_0=0} 
{\rm Exp}[-m_0/0.1\,{\rm eV}]$. This helps us in finding the required perturbations of $\delta M_\nu$ as a function of experimentally determined 
quantities such as solar and atmospheric mass squared differences and the mixing angles. To this end we apply the result of bottom-up approach 
to develop a model which predict large $\theta_{13}$ perturbatively. 

The paper is organized as follows. In section~\ref{bua}, we start with a brief description of the bottom-up 
approach that is developed in this paper to find the desired perturbation of the mass matrix in the 
flavor basis. Then, in section~\ref{tbm}, we use the knowledge of our current experimental results
to find the deviation in the tree-level mass matrix of the neutrinos, obtained by using TBM mixing ansatz. 
In section~\ref{model}, we use the result of bottom-up approach to develop a model in which the perturbation to the 
TBM mixing is obtained. We then fit the model parameters using the data from the bottom-up approach of section~\ref{tbm} 
and conclude in section~\ref{con}.

\section{Phenomenology}
\label{pheno}
\subsection{Bottom up approach}
\label{bua}
The recent global fit to the neutrino oscillation data has ruled out the possibility of a zero reactor angle at $10.2\sigma$
confidence level~\cite{Tortola,Schwetz}. Evidence of non zero $\theta_{13}$ at $3\sigma$ level was first established by data 
from T2K~\cite{t2k}, MINOS~\cite{minos} and Double Chooz~\cite{DC}. More recently, Daya Bay~\cite{daya} and RENO~\cite{reno} experiments 
confirm large $\theta_{13}$ at more than $5\sigma$ confidence level~(CL) from the reactor $\bar{\nu}_e \to \bar{\nu}_e$ oscillations. 
The current best fit values and the $3\sigma$ allowed values of all the mixing parameters are summarized in Table.~\ref{bestfit}. 
\begin{table}[ht]
\centering                          
\begin{tabular}{|c|c|c|}           
\hline\hline                      
Oscillation parameters & Best fit value & $3\sigma$ range \\ [0.5ex]   
\hline                              
$\Delta m_{\odot}^2\left[ 10^{-5}\,{\rm eV^2} \right]$ & $7.62$  & $[7.12 - 8.20]$ \\               
$|\Delta m_{\rm atm}^2|\left[ 10^{-3}\,{\rm eV^2} \right] $ & $2.55~(2.43)$  & $[2.31 - 2.74]~([2.21 - 2.64])$ \\
$\sin^2\,\theta_{12}$ & $0.320$ & $[0.27 - 0.37]$ \\
$\sin^2\,\theta_{23}$ & $0.613~(0.600)$ & $[0.36 - 0.68]~([0.37 - 0.67])$  \\
$\sin^2\,\theta_{13}$ & $0.0246~(0.0250)$ & $[0.017 - 0.033]$ \\ [1ex]         
\hline                              
\end{tabular}
\caption{Current status of oscillation parameters taken from Ref.~\cite{Tortola}.}   
\label{bestfit}          
\end{table}
The values written within brackets are for inverted ordering of the neutrino mass spectrum. At this juncture we note 
that the only unknown quantity in the neutrino mass matrix is the lightest neutrino mass apart from the CP violating 
phases (one Dirac and two Majorana phases).

The neutrino mixing matrix derived from the current best fit parameters is
\begin{eqnarray}
\label{uexpt}
&&U^c_{{\rm EXPT}} = \left( \begin{array}{ccc}
 0.81442 & 0.55868 & 0.15684 \\
-0.45317 & 0.44353 & 0.77325 \\
 0.36244 & -0.70083 & 0.61439 \end{array} \right),
\end{eqnarray}
and the corresponding $3\sigma$ allowed values of the mixing matrix is
\begin{eqnarray}
\label{uexpt3sig}
U_{{\rm EXPT}} = \left( \begin{array}{ccc}
0.78 - 0.85 & 0.51 - 0.60  & 0.13 - 0.18 \\
-(0.39 - 0.57) & 0.36 - 0.64 & 0.59 - 0.82 \\
0.19 - 0.44 & -(0.54 - 0.76) & 0.56 - 0.79 \end{array} \right)\,,
\end{eqnarray}
where we have assumed $\delta_{CP}$ to be zero for simplicity. The corresponding neutrino mass matrix in the flavor basis can be written as
\begin{eqnarray}\label{mexpt}
(M_{\nu})_{{\rm EXPT}} = U_{{\rm EXPT}}\,{\rm diag}(m_1,\,m_2,\,m_3)\,U^{T}_{{\rm EXPT}} 
\end{eqnarray}
where $(M_{\nu})_{{\rm EXPT}}$ is a real symmetric matrix. Therefore, it is described by three masses: $m_1,\,m_2,\,m_3$ and 
three mixing angles: $\theta_{12},\,\theta_{23},\,\theta_{13}$. The mass eigenvalues $m_1,\,m_2,\,m_3$ in the basis where the charged 
leptons are real and diagonal can be expressed as:
$m_2 = \sqrt{\Delta m_{\odot}^2 + m_1^2}$ and $m_3 = \sqrt{\Delta m_{\rm atm}^2 + m_1^2}$ for normal hierarchy and
$m_1 = \sqrt{m_3^2 + \Delta m_{\rm atm}^2}$ and $m_2 = \sqrt{m_3^2 + \Delta m_{\rm atm}^2 + \Delta m_{\odot}^2}$ for inverted hierarchy.   

We define the perturbed neutrino mass matrix $\delta M_\nu$ as 
\begin{eqnarray}
\delta M_\nu &=&
(M_{\nu})_{{\rm EXPT}} - (M_{\nu})_0 \,,
\end{eqnarray}
where $(M_{\nu})_{{\rm EXPT}}$ is the neutrino mass matrix given by Eq.~\ref{mexpt} and $(M_{\nu})_0$ is the neutrino mass matrix 
corresponding to a mixing matrix $U_0$, that is obtained in a specific model such as TBM mixing, BM mixing 
and/or DC mixing {\it etc}. This $\delta M_\nu$ is a symmetric matrix whose elements denote the amount of perturbation 
that is needed to add to the $(M_{\nu})_0$ so that it is consistent within the $3\sigma$ of the experimental results. In order 
to gauge the size of the $\delta M_\nu$ matrix, we do a random scan of all the parameters: $\theta_{13}$, $\theta_{23}$, $\theta_{12}$, 
$\Delta m_{\odot}^2$ and $|\Delta m_{\rm atm}^2|$ within their $3\sigma$ allowed range of values. Note that $(M_{\nu})_0$ depends only on 
the value of the lightest neutrino mass $m_0$ as the mixing angles are expected to be fixed via certain symmetries as done in TBM, BM 
and/or DC mixing scenarios. Hence, to see the effect of $m_0$ on the elements of $\delta M_\nu$ we vary $m_0$ in the range 
$[10^{-9}, 10]\,{\rm eV}$. We then define the predicted neutrino mass matrix $(M_{\nu})_M$ of any model as:
\begin{eqnarray}
(M_{\nu})_M &=&
(M_{\nu})_0 + \delta M_\nu\,.
\end{eqnarray}
Note that the matrix elements of $(M_{\nu})_M$ and $\delta M_\nu$ are function of the oscillation parameters $\theta_{12}$,
$\theta_{23}$, and $\theta_{13}$ as well as of the lightest neutrino mass $m_0$. We then perform a naive statistical analysis and 
compare the elements of the $(M_{\nu})_M$ to the experimental data of Eq.~\ref{mexpt} by a $\chi^2$ function which is
defined by:
\begin{eqnarray}
\chi^2 &=&
\Big[(M_{\nu}^c)_{{\rm EXPT}} - (M_{\nu})_M \Big]^2/m_0\sigma^2 = \Big[\delta M_\nu^c - \delta M_\nu\Big]^2/m_0\sigma^2\,,
\end{eqnarray}
where $\sigma = \sqrt{(M_{\nu}^c)_{{\rm EXPT}}}$ and $\delta M_\nu^c = (M_{\nu}^c)_{{\rm EXPT}} - (M_{\nu})_0$.
Note that $(M_{\nu}^c)_{{\rm EXPT}}$ is the experimental mass matrix corresponding to the best fit values of
the oscillation parameters and is diagonalized by the mixing matrix of Eq.~\ref{uexpt}. It is worth noting that as 
$\delta M_\nu \to \delta M_\nu^c$, we get the minimum $\chi^2$ and hence $\delta M_\nu^c$ is the required amount of 
perturbation that should be added to $(M_{\nu})_0$. We will discuss the properties of the $\delta M_\nu$ matrix using 
the TBM mixing ansatz. Although we focus mainly on the TBM mixing, some of the features presented here are 
applicable to other tree-level mass matrix.
 
\subsection{Perturbation in Tribimaximal mixing Scenario}
\label{tbm}
We now proceed to discuss the tribimaximal (TBM) mixing scenario. For the TBM mixing , we have $s_{12} = 1/\sqrt{3}$, 
$s_{23} = 1/\sqrt{2}$, and $s_{13} = 0$. We know that apart from the recently measured value of $\theta_{13}$, the solar 
and atmospheric mixing angles are in a very good agreement with the TBM ansatz. Modifications to this TBM 
scenario have been studied by many authors~\cite{tbmprtb}.

We begin by writing the TBM mixing matrix in the standard parameterization of Eq.~\ref{upmns}, that is
\begin{eqnarray}
\label{utbm}
U_{{\rm TBM}} &=& \left( \begin{array}{ccc}
\sqrt{2/3} & \sqrt{1/3}  & 0 \\
-\sqrt{1/6} & \sqrt{1/3} & \sqrt{1/2} \\
\sqrt{1/6} & -\sqrt{1/3} & \sqrt{1/2} \end{array} \right)
=
\left( \begin{array}{ccc}
0.81650 & 0.57735  & 0 \\
-0.40825 & 0.57735 & 0.70711 \\
0.40825 & -0.57735 & 0.70711 \end{array} \right).
\end{eqnarray}

\begin{figure}[htbp]
\begin{center}
\includegraphics[width=8cm,height=5cm]{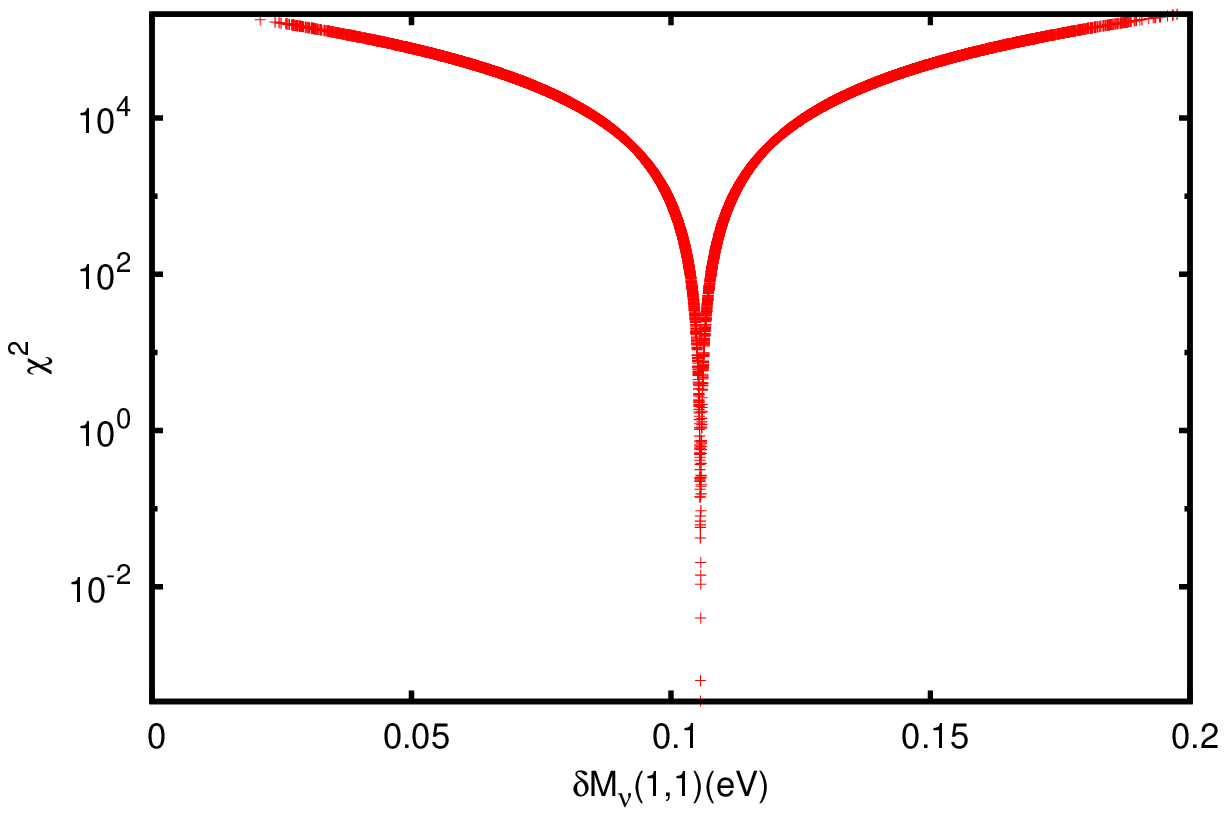}
\includegraphics[width=8cm,height=5cm]{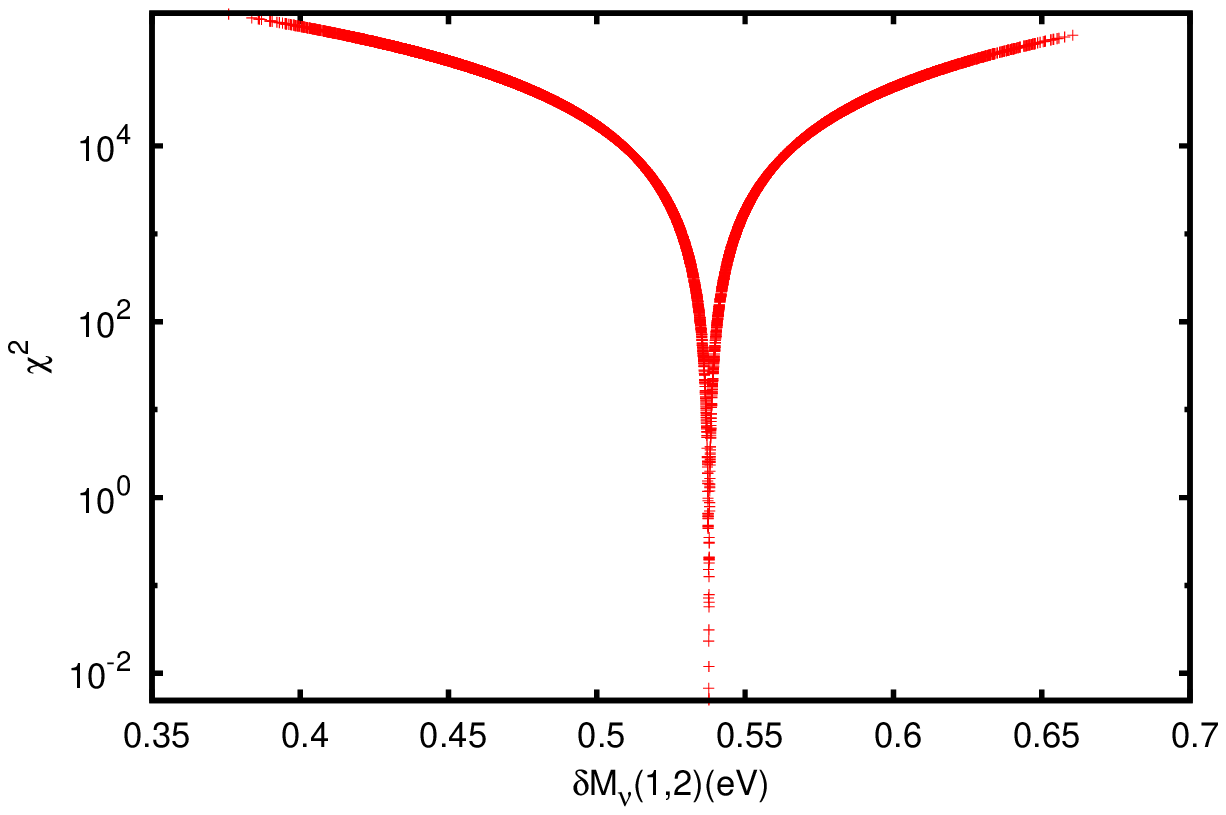}
\includegraphics[width=8cm,height=5cm]{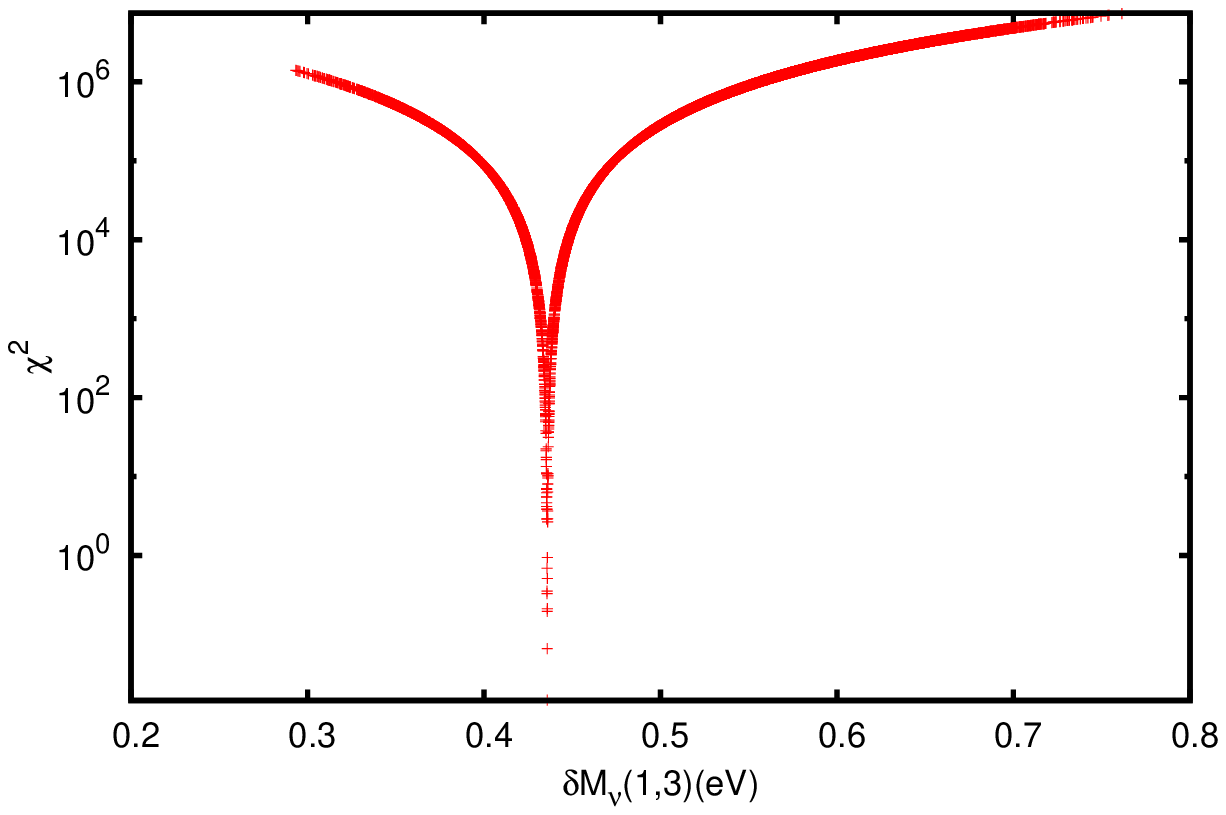}
\includegraphics[width=8cm,height=5cm]{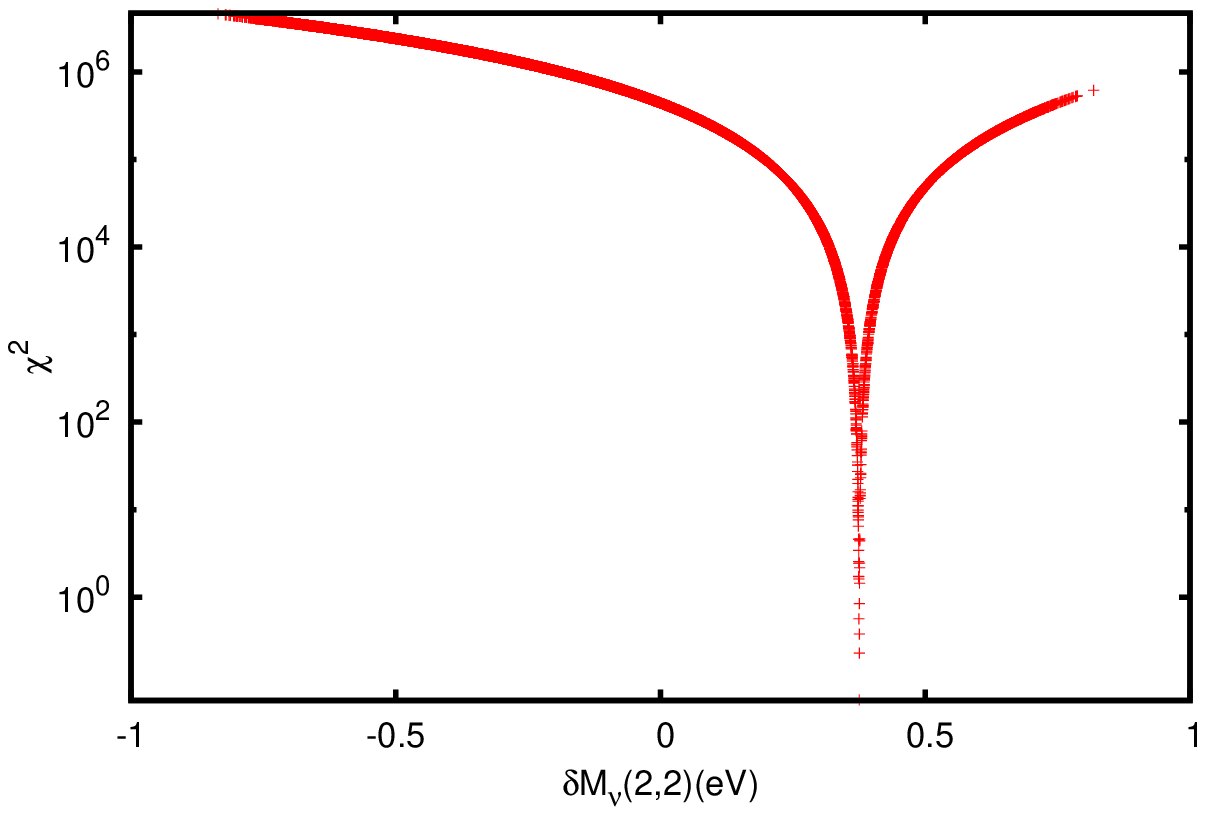}
\includegraphics[width=8cm,height=5cm]{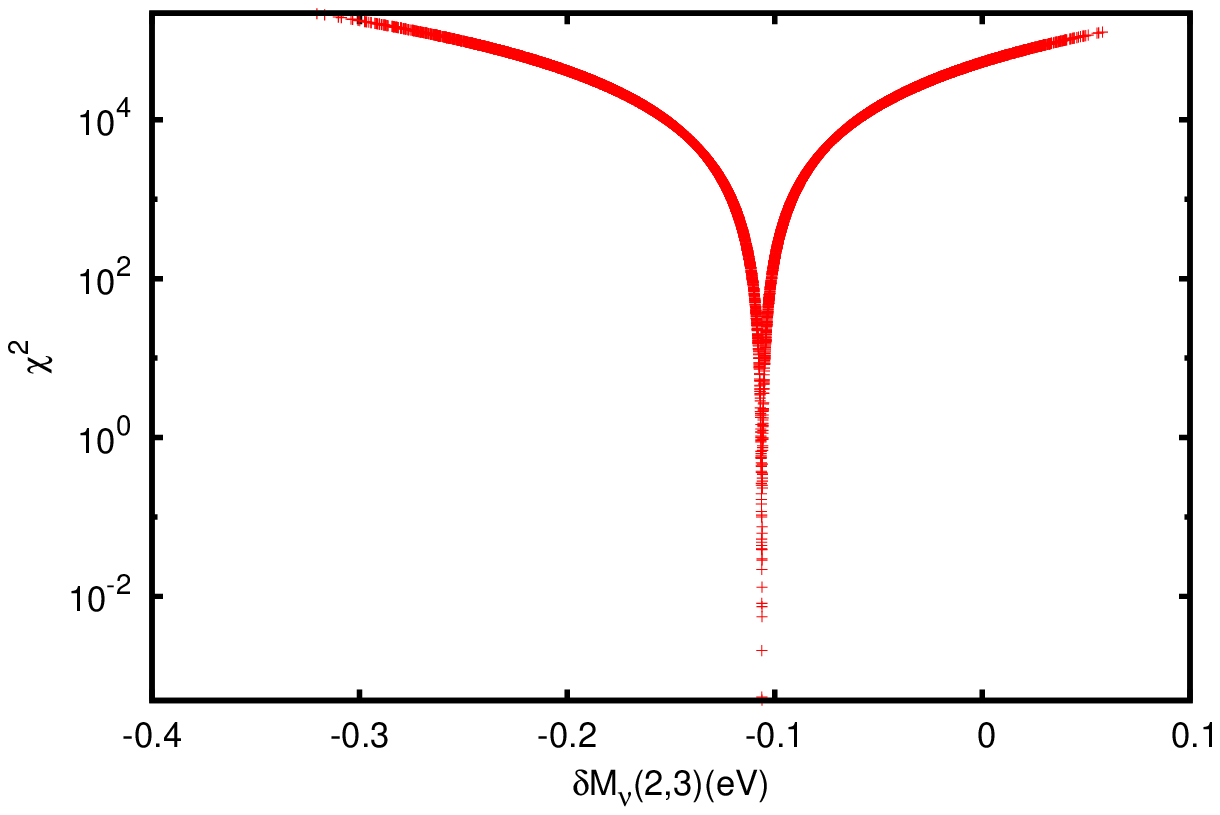}
\includegraphics[width=8cm,height=5cm]{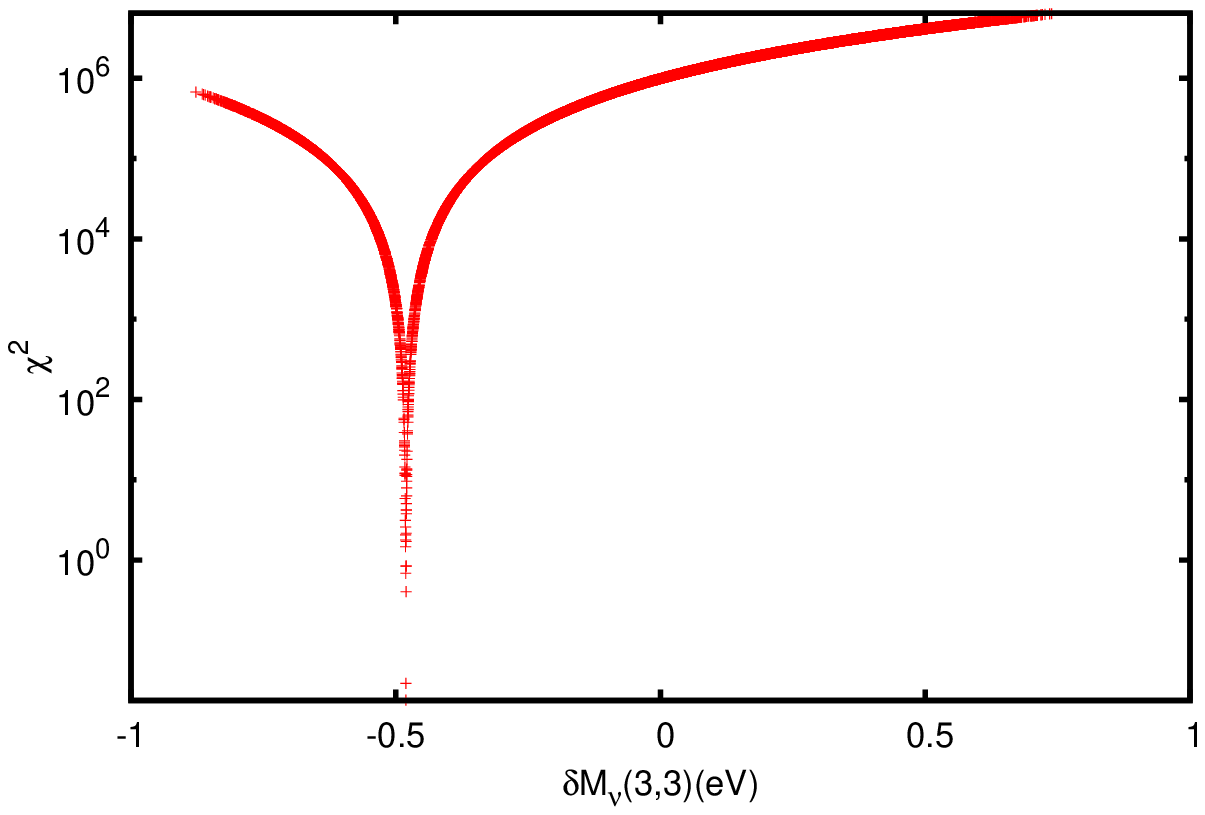}
\end{center}
\caption{$\chi^2$ of $(\delta M_\nu (i,j)\times 10^2)$ for normal ordering of neutrino masses with TBM mixing in 
tree-level.}
\label{chisqep_tbm}
\end{figure}

\begin{figure}[htbp]
\begin{center}
\includegraphics[width=8cm,height=5cm]{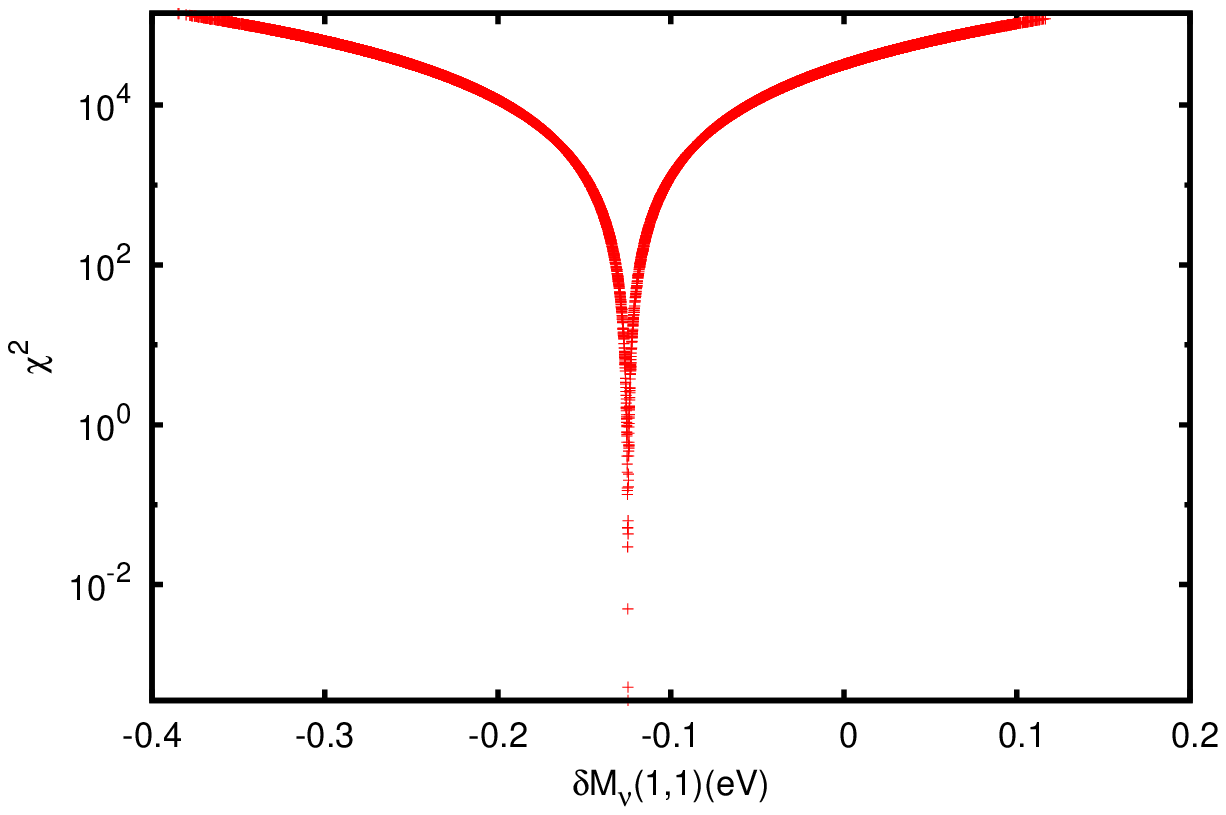}
\includegraphics[width=8cm,height=5cm]{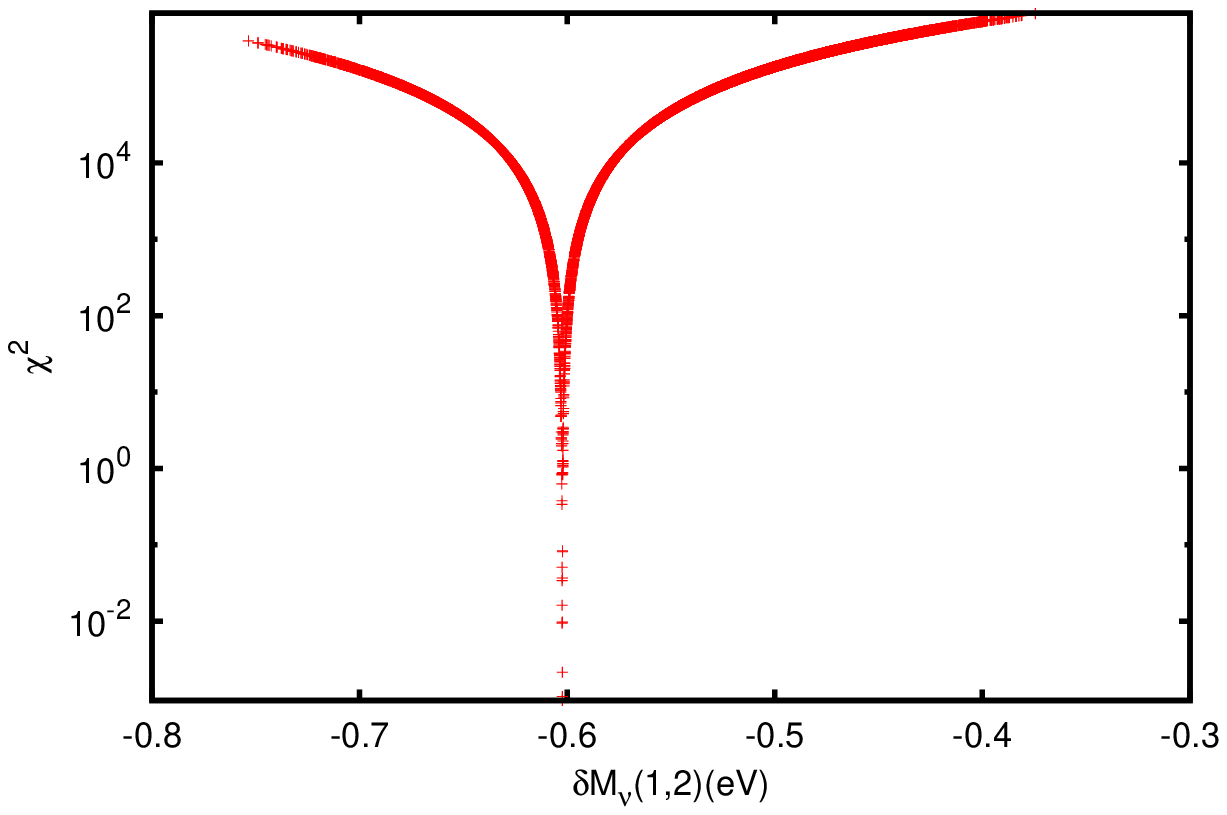}
\includegraphics[width=8cm,height=5cm]{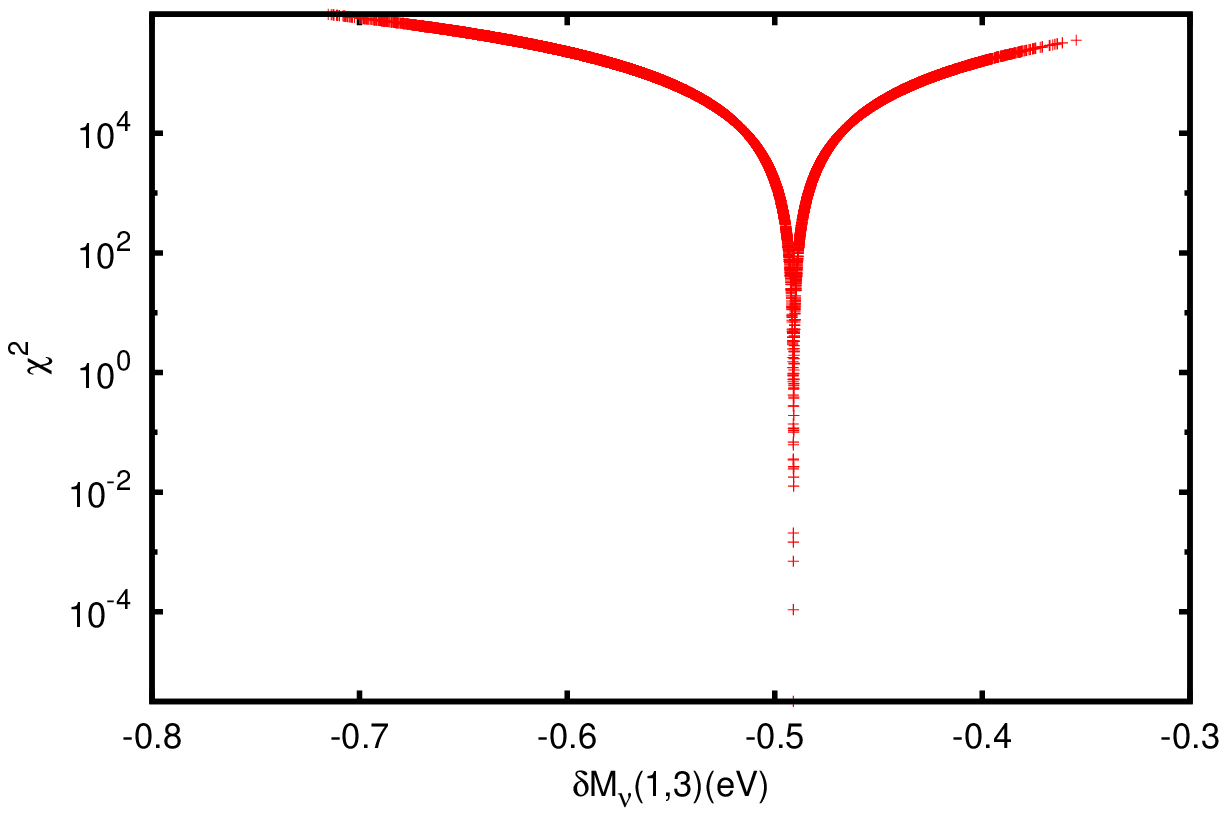}
\includegraphics[width=8cm,height=5cm]{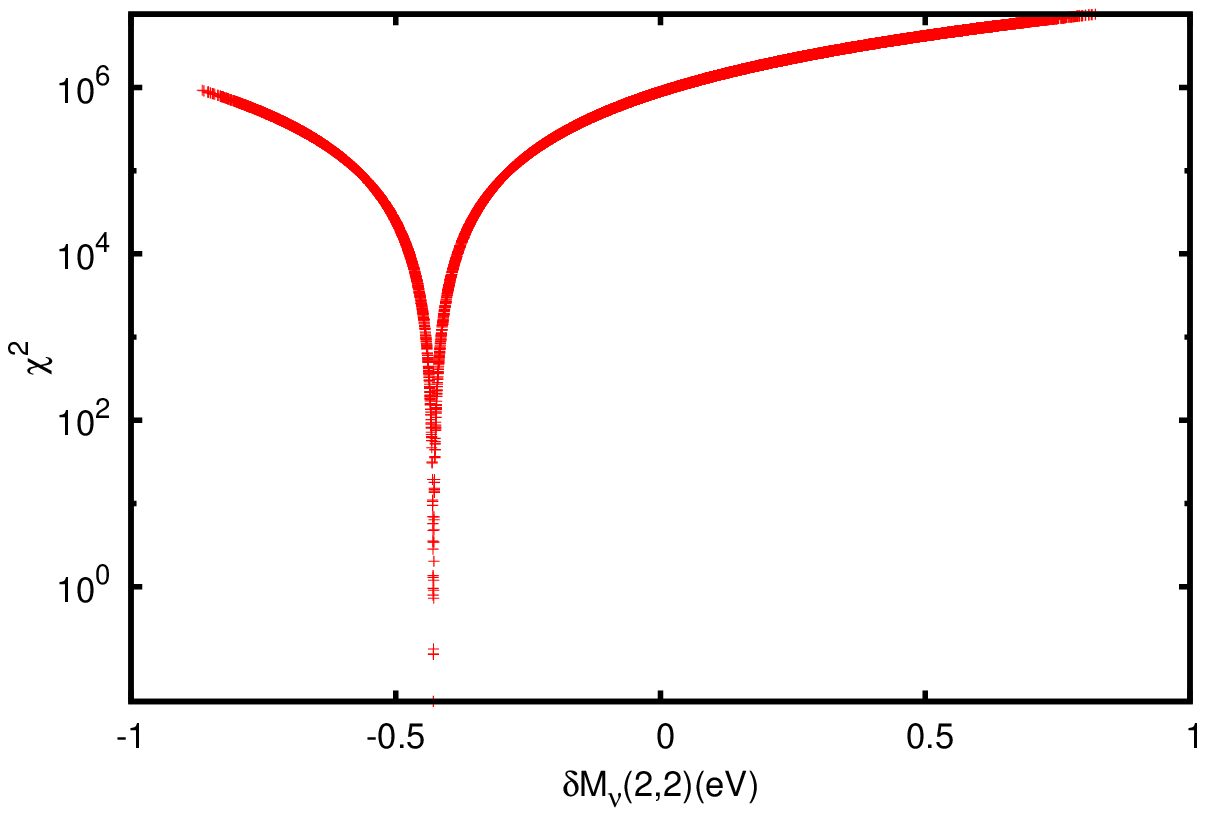}
\includegraphics[width=8cm,height=5cm]{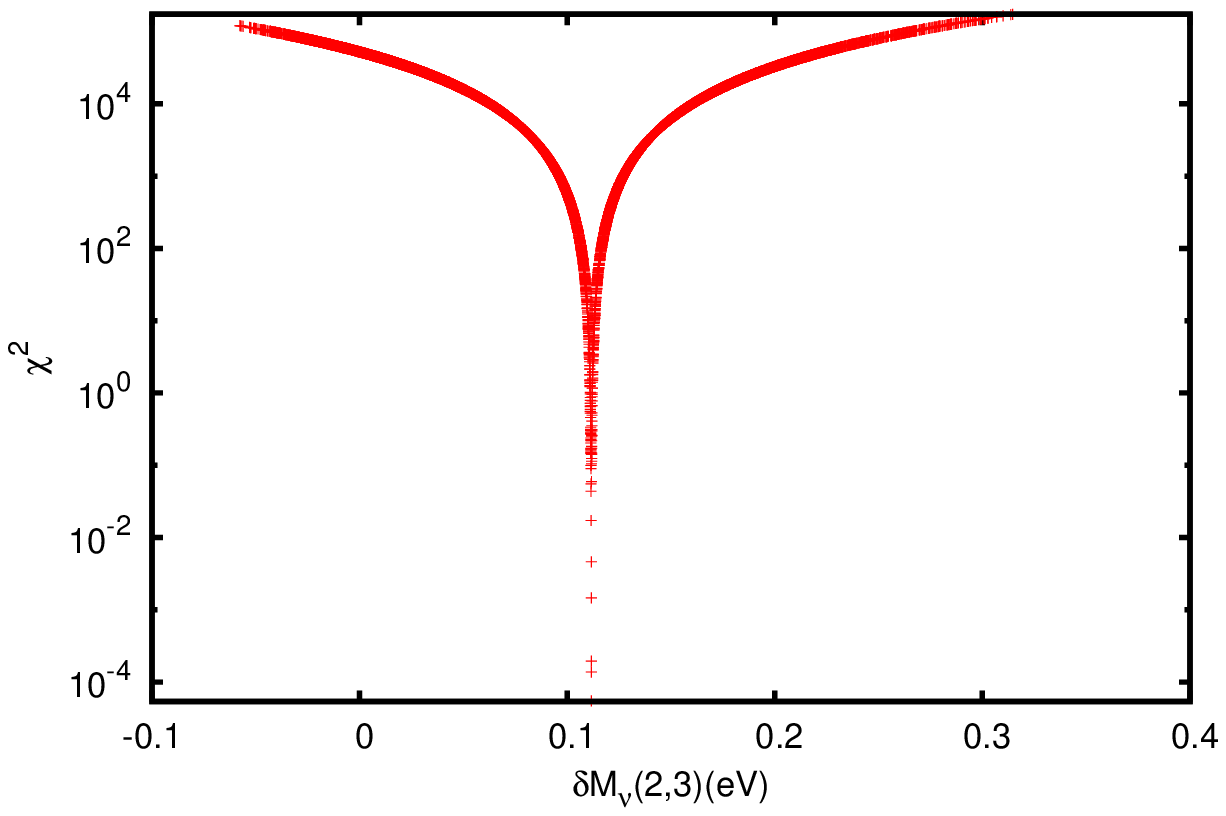}
\includegraphics[width=8cm,height=5cm]{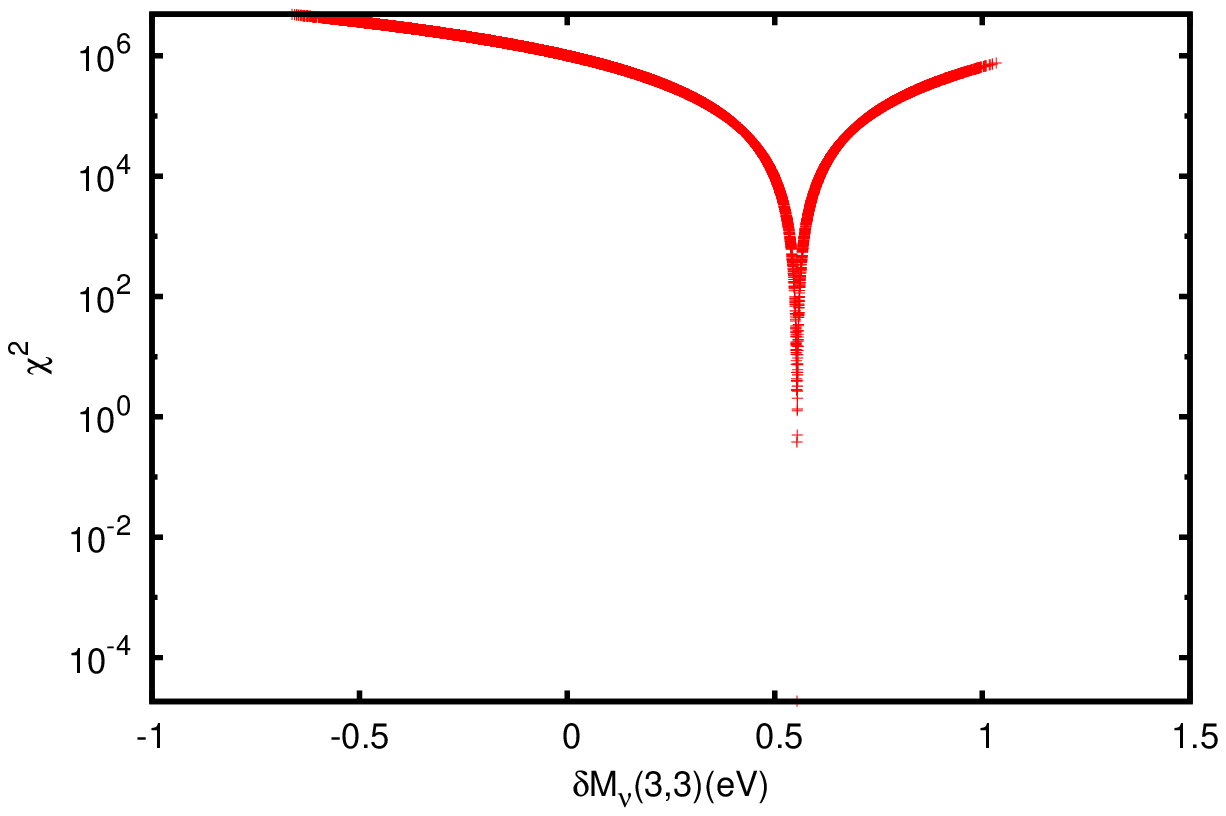}
\end{center}
\caption{$\chi^2$ of $(\delta M_\nu (i,j)\times 10^2)$ for inverted ordering of neutrino masses with TBM mixing in tree-level.}
\label{chisqep_tbm_ih}
\end{figure}

\begin{figure}[htbp]
\begin{center}
\includegraphics[width=8cm,height=5cm]{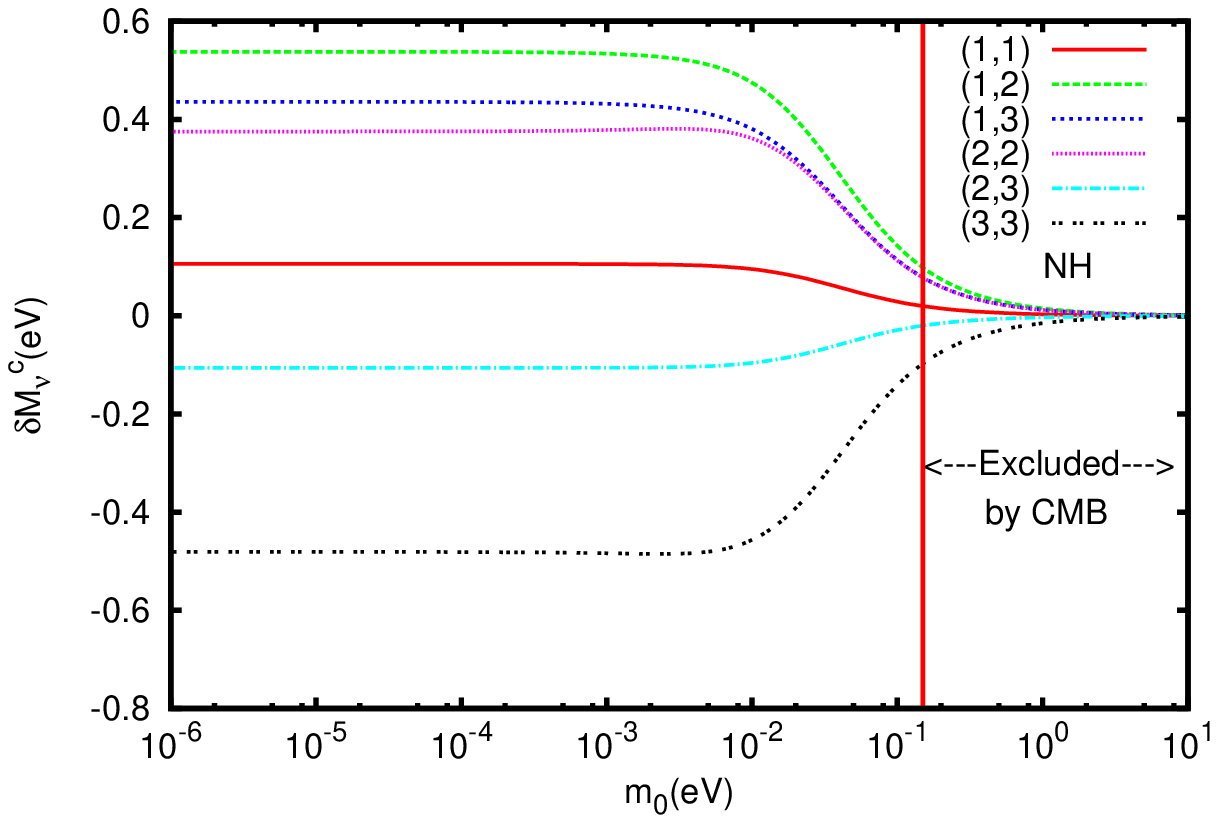}
\includegraphics[width=8cm, height=5cm]{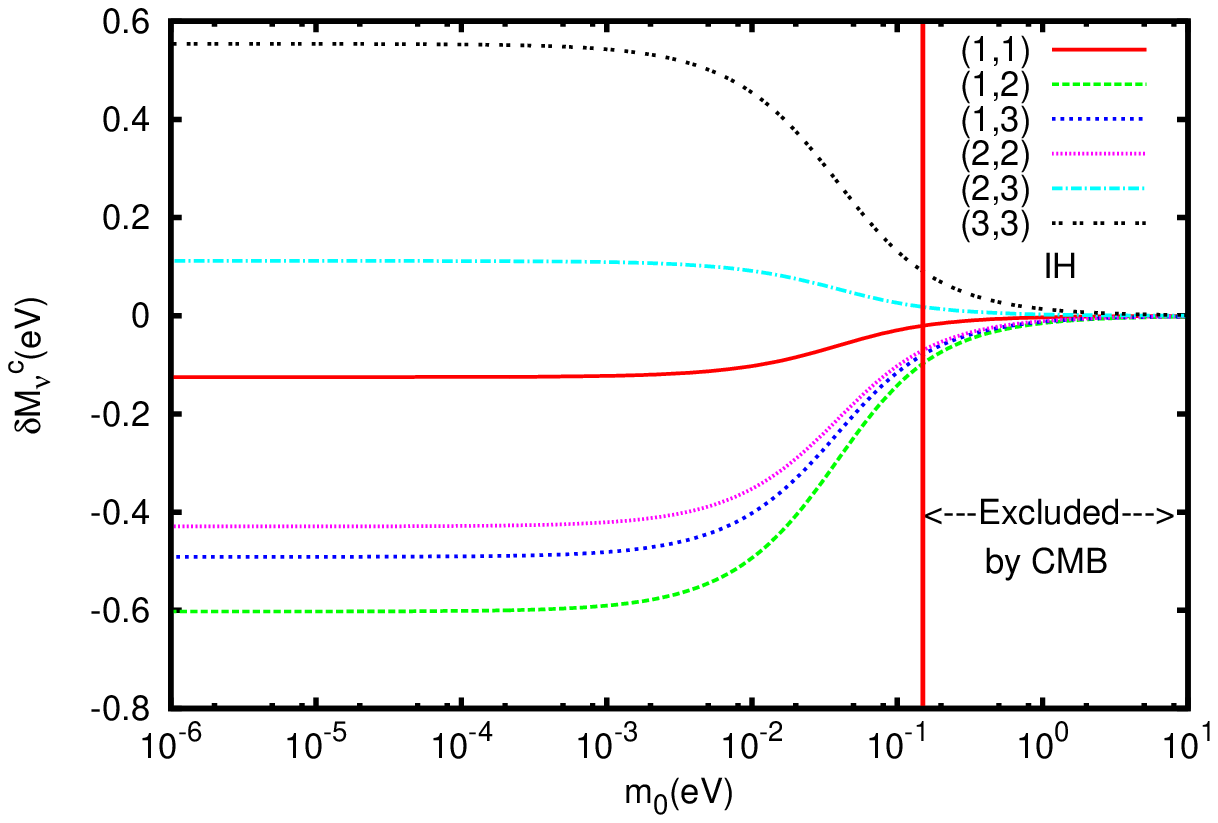}
\end{center}
\caption{Variation of $(\delta M_\nu^c \times 10^2)$ with respect to the lightest neutrino mass for NH (left) and IH (right) of neutrino masses with TBM mixing. An 
upper cut-off at $m_0 \approx 0.15 {\rm eV}$~\cite{wmap9} is put to ensure that $\delta M_\nu (i,j)\neq 0$ for all $m_0 < 0.15 {\rm eV}$.  
}
\label{epm_tbm}
\end{figure}

The corresponding neutrino mass matrix in the flavor basis and the perturbed matrix $\delta M_\nu$ are
\begin{eqnarray}
\label{mtbm}
(M_{\nu})_{{\rm TBM}} &=& U_{{\rm TBM}}\,{\rm diag}(m_1,\,m_2,\,m_3)\,U^{T}_{{\rm TBM}}\,, \qquad\qquad
\delta M_\nu = (M_{\nu})_{{\rm EXPT}} - (M_{\nu})_{{\rm TBM}}.
\end{eqnarray}
We wish to determine the elements of this $\delta M_\nu$ matrix in a bottom-up approach. We follow the procedure 
given in section~\ref{bua} and perform a random scan over all the oscillation parameters within their $3\sigma$ 
range of values. In Fig.~\ref{chisqep_tbm}, we show $\delta M_\nu (i,j)$ versus $\chi^2$ for $m_0 = 10^{-9}\,{\rm eV}$. 
The minimum of each curve corresponds to the value of $\delta M_\nu^c(i,j)$. A similar plot is shown in Fig.~\ref{chisqep_tbm_ih} 
for inverted hierarchy as well.
We know that the neutrino mass matrix in the flavor basis does depend on the value of the lightest mass eigenvalue  $m_0$.
In order to see the effect of $m_0$ on the value of $\delta M_\nu^c(i,j)$ qualitatively, we vary
$m_0$ in the range $(10^{-10}$, $10)\,{\rm eV}$ while keeping other parameters such as $\theta_{12}$, $\theta_{13}$, $\theta_{23}$,
$|\Delta m_{\rm atm}^2|$, and $\Delta m_{\odot}^2$ fixed at their best fit values. In Fig.~\ref{epm_tbm}, we plot $\delta M_\nu^c(i,j)$ versus 
$m_0$ and it is clear that for $m_0 < 10^{-3}\,{\rm eV}$, the value of $\delta M_\nu^c(i,j)$ remains constant. In other words, 
$\delta M_\nu(i,j)$ are independent of the value of $m_0$. However, for $m_0 \gsim 10^{-3}\,{\rm eV}$, the value
of $\delta M_\nu(i,j)$ keeps changing. In the limit, $m_0 >> \sqrt{|\Delta m_{\rm atm}^2|} = 4.9\times 10^{-2}\,{\rm eV}$, we have 
$m_3=m_2=m_1=m_0$ and hence $\delta M_\nu (i,j)$ become zero. This particular feature is common to both normal and inverted 
hierarchies as can be seen from Fig.~\ref{epm_tbm}. 
To appreciate the plots in Fig.~\ref{epm_tbm}, we write down $\delta M_\nu(i,j)$ as a function of neutrino masses and 
mixing angles:  
\begin{eqnarray}
 \delta M_\nu (1,1)  &=&  ({(c_{12}c_{13})}^2-2/3)m_1 + ({(s_{12}c_{13})}^2-1/3)m_2 + s^2_{13}m_3 \nonumber\\
\delta M_\nu (1,2)   &=& (c_{12}c_{13}(-s_{12}c_{23}-c_{12}s_{23}s_{13})+1/3)m_1\nonumber\\
                     &+& (s_{12}c_{13}(c_{12}c_{23}- s_{12}s_{23}s_{13})-1/3)m_2 + s_{13}c_{13}s_{23}m_3 \nonumber\\
\delta M_\nu(1,3)    &=& (c_{12}c_{13}(s_{12}s_{23}-c_{12}c_{23}s_{13})-1/3)m_1 \nonumber\\
                     &+& (s_{12}c_{13}(-c_{12}s_{23}-s_{12}c_{23}s_{13}) +1/3)m_2 + s_{13}c_{13}c_{23}m_3 \nonumber\\
\delta M_\nu (2,2)  &=& ({(-s_{12}c_{23}-c_{12}s_{23}s_{13})}^2-1/6)m_1\nonumber\\
                    &+& ({(c_{12}c_{23}-s_{12}s_{23}s_{13})}^2-1/3)m_2 + ({(c_{13}s_{23})}^2-1/2)m_3 \nonumber\\
\delta M_\nu (2,3)  &=& ((-s_{12}c_{23}-c_{12}s_{23}s_{13})(s_{12}s_{23}-c_{12}c_{23}s_{13})+ 1/6)m_1 \nonumber\\
                    &+& ((c_{12}c_{23}-s_{12}s_{23}s_{13})(-c_{12}s_{23}-s_{12}c_{23}s_{13})+1/3)m_2 + (s_{23}c_{13}c_{23}c_{13}-1/2)m_3\nonumber\\
\delta M_\nu (3,3)  &=& ({(s_{12}s_{23}-c_{12}c_{23}s_{13})}^2-1/6)m_1 \nonumber\\
                    &+& ({(-c_{12}s_{23}-s_{12}c_{23}s_{13})}^2-1/3)m_2 + ({(c_{13}c_{23})}^2-1/2)m_3\,.
\label{delta_mu_nu}
\end{eqnarray} 
Note that the higher eigenvalues $m_2$ and $m_3 (m_1)$ can be re-expressed as a function of lightest neutrino mass $m_0\equiv m_1 (m_3)$. Therefore, 
it is obvious that all the elements: $\delta M_\nu (i,j)$ are function of the only unknown quantity $m_0$. From Fig.~\ref{epm_tbm} 
we see that $\delta M_\nu(1,1)$, $\delta M_\nu (1,2)$, $\delta M_\nu(1,3)$, $\delta M_\nu(2,2)$ start with a positive value, while $\delta M_\nu (2,3)$ and 
$\delta M_\nu (3,3)$ start with a negative value. An exactly opposite spectrum is observed in case of inverted hierarchy as expected.
In either case we observe that for $m_0 << 10^{-3} {\rm eV}$, $\delta M_\nu (i,j)$ are independent of $m_0$. Therefore, in this limit it is reasonable 
to set $m_0 \approx 0$.

In the opposite limit when $m_0 >> \sqrt{|\Delta m_{\rm atm}^2|} = 4.9\times 10^{-2}\,{\rm eV}$, we have $m_3=m_2=m_1=m_0$. In this limit we get 
$M_\nu(i,j) = m_0\,(I) $, where $I$ is the identity matrix. Hence all the $\delta M_\nu(i,j)$ are zero as expected~\footnote{ In our case we set all 
the CP-violating phases to be zero. As a result, the unitary PMNS matrix is simply an orthogonal matrix and hence we get $M_\nu(i,j) = {\cal O} m_0\,(I)
{\cal O}^T =m_0\,(I) $. This is not true if one assumes the CP-violating phases to be non-zero.}. This feature can be easily read from Fig.~\ref{epm_tbm}. 
However, the observation tells us that this should not be the case and we need small mass splittings between the mass eigenvalues in order to satisfy the 
solar and atmospheric mass differences. Therefore, we put an upper cut-off at $m_0 \approx 0.15 {\rm eV}$~\cite{wmap9}, as required by CMB data, to ensure that 
$\delta M_\nu (i,j)\neq 0$ for all $m_0 < 0.15 {\rm eV}$. In this region $m_1 \neq m_2 \neq m_3$.  

For $10^{-3}\,{\rm eV} < m_0 < 0.15 {\rm eV}$, which is comparable to solar and atmospheric values, we observe appreciable effect of $m_0$ 
on various $\delta M_\nu(i,j)$ as shown in Fig.~\ref{epm_tbm}. We have factored out this dependency of all $\delta M_\nu(i,j)$ on 
$m_0$ using an exponential parameterization:   
\begin{eqnarray}\label{m_0}
\left|\delta M_\nu (m_0,\,\sqrt{\Delta m^2_{\odot}},\, \sqrt{|\Delta m^2_{\rm atm}|},\,\theta_{12},\,\theta_{13},\,\theta_{23})\right|  &=& 
\left|\delta M_\nu (m_0 = 0,\,\sqrt{\Delta m^2_{\odot}},\, \sqrt{|\Delta m^2_{\rm atm}|},\,\theta_{12},\,\theta_{13},\,\theta_{23})\right|\nonumber\\
&& \times {\rm Exp}(-m_0/a)\,.
\end{eqnarray}
where $a$ is determined from a fit to the recent experimental data and found to be $a \approx 0.1\,{\rm eV}$. As a result 
we could do all our analysis for $m_0 \le 10^{-3}\,{\rm eV}$ which is equivalent to setting $m_0=0$. Then we generalize it to any value of 
$m_0$ using Eq.~\ref{m_0}. 

We first compare the perturbed matrix elements $\delta M_\nu^c(i,j)$ with the elements of tree level mass 
matrix $(M_{\nu})_{{\rm TBM}}(i,j)$. In case of normal hierarchy, the experimental mass matrix, the tree-level neutrino mass matrix,
and its deviations from the central values are given by:  
\begin{eqnarray}
&&\frac{    (M_{\nu}^c)_{{\rm EXPT}}   } {   10^{-2}\,{\rm eV}     }=
\left( \begin{array}{ccc}
0.397 & 0.829  & 0.145 \\
0.829 & 3.191 & 2.128 \\
0.145 & 2.128 & 2.335 \end{array} \right),\qquad\qquad
\frac{      (M_{\nu})_{{\rm TBM}}    }{   10^{-2}\,{\rm eV}     }=
\left( \begin{array}{ccc}
0.291 & 0.291  & -0.291 \\
0.291 & 2.816 & 2.234 \\
-0.291 & 2.234 & 2.816 \end{array} \right)\,,
\label{mnu_expt_tbm}
\end{eqnarray}
and
\begin{eqnarray}
\frac{\delta M_\nu^c}{10^{-2}\,{\rm eV}} = \left( \begin{array}{ccc}
0.106 & 0.538  & 0.436 \\
0.538 & 0.375 & -0.106 \\
0.436 & -0.106 & -0.481 \end{array} \right)\,.
\label{NH_perturbation}
\end{eqnarray}
Similarly, for inverted hierarchy, the mass matrix $(M_{\nu}^c)_{{\rm EXPT}}$ and $(M_{\nu})_{{\rm TBM}}$ are,
\begin{eqnarray}
&& \frac{(M_{\nu}^c)_{{\rm EXPT}}}{10^{-2}\,{\rm eV}}=
\left( \begin{array}{ccc}
4.830 & -0.577  & -0.517 \\
-0.577 & 2.061 & -2.379 \\
-0.517 & -2.379 & 3.044 \end{array} \right), \qquad
 \frac{    (M_{\nu})_{{\rm TBM}}  }{   10^{-2}\,{\rm eV}        } =
\left( \begin{array}{ccc}
4.956 & 0.026 & -0.026 \\
0.026 & 2.490 & -2.490 \\
-0.026 & -2.490 & 2.490 \end{array} \right)
\end{eqnarray}
and the corresponding perturbed matrix is
\begin{eqnarray}
\frac{\delta M_{\nu}^c}{10^{-2}\,{\rm eV}} = \left( \begin{array}{ccc}
-0.125 & -0.602  & -0.491 \\
-0.602 & -0.429 & 0.112 \\
-0.491 & 0.112 & 0.554 \end{array} \right).
\end{eqnarray}
We introduce a new parameter $\epsilon$ as:
\begin{equation}
\epsilon^c(i,j)=\frac{\delta M_\nu^c(i,j)}{(M_{\nu})_{{\rm TBM}}(i,j)}
\end{equation}
such that it will give a measure of required perturbation with respect to the corresponding tree-level value. Thus for normal and 
inverted hierarchies of neutrino mass spectrum, we get:
\begin{eqnarray}
&&(\epsilon^c)_{\rm NH} =\left(
 \begin{array}{ccc}
0.364 & 1.848  & -1.848 \\
1.848 & 0.133 & -0.047 \\
-1.848 & -0.047 & 0.171 \end{array} \right)  \qquad \qquad
(\epsilon^c)_{\rm IH} = \left(
\begin{array}{ccc}
0.02 & 23.154  & 18.885 \\
23.154 & 0.172 & 0.045 \\
18.885 & 0.045 & 0.222 \end{array} \right)\,. 
\label{epsilon}
\end{eqnarray}
From Eq.~(\ref{epsilon}), it is clear that $(1,2)$, and $(1,3)$ elements of the TBM mass matrix needs 
to be modified largely to be consistent with the experiment. The perturbations of rest of the elements are 
small in comparison to their tree-level masses. In case of NH the modification is mild, while it 
is significant in case of IH case. 

\begin{figure}[htbp]
\begin{center}
\includegraphics[width=8cm,height=5cm]{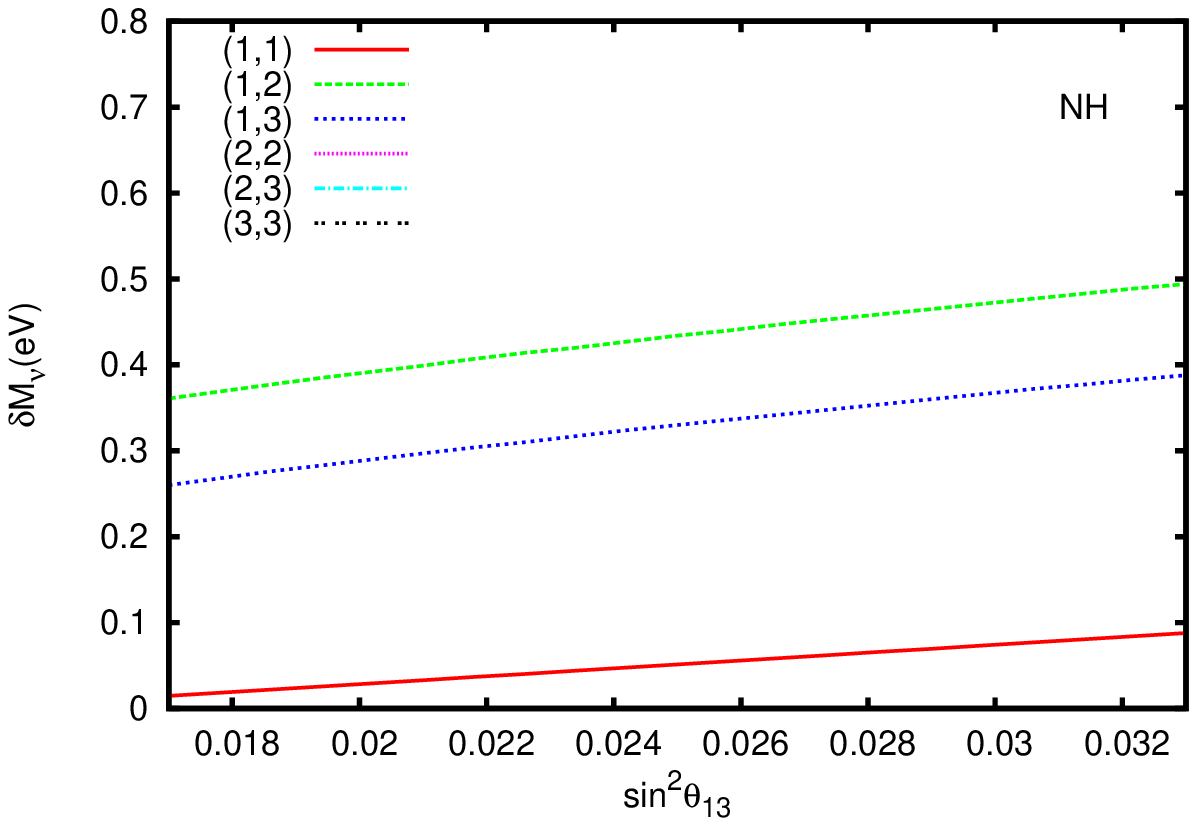}
\includegraphics[width=8cm, height=5cm]{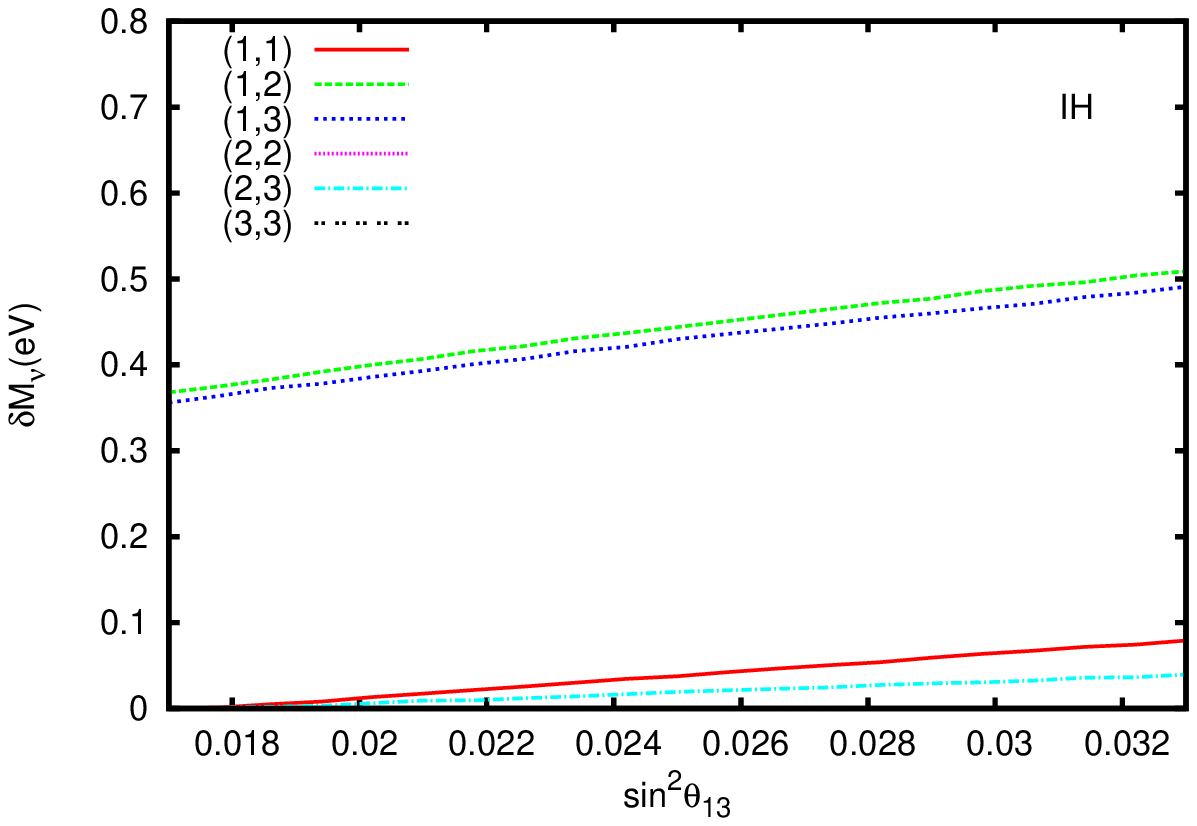}
\includegraphics[width=8cm,height=5cm]{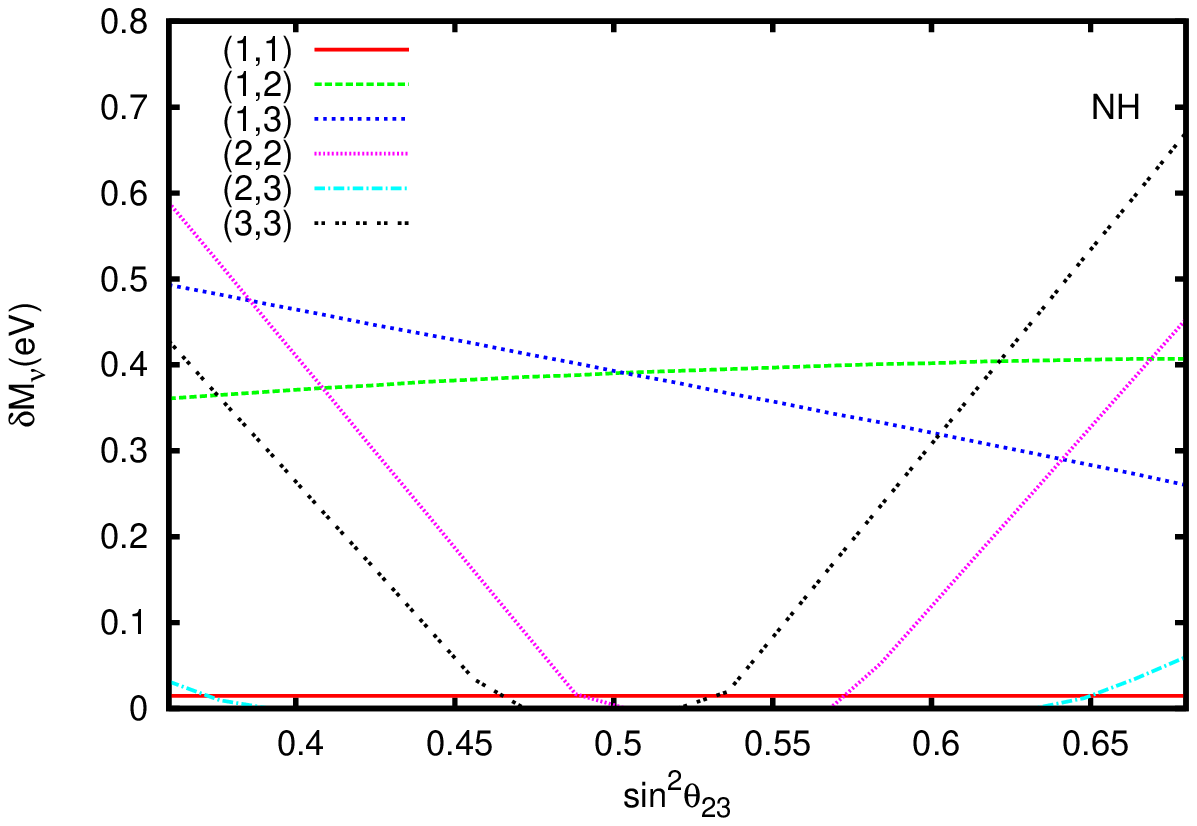}
\includegraphics[width=8cm, height=5cm]{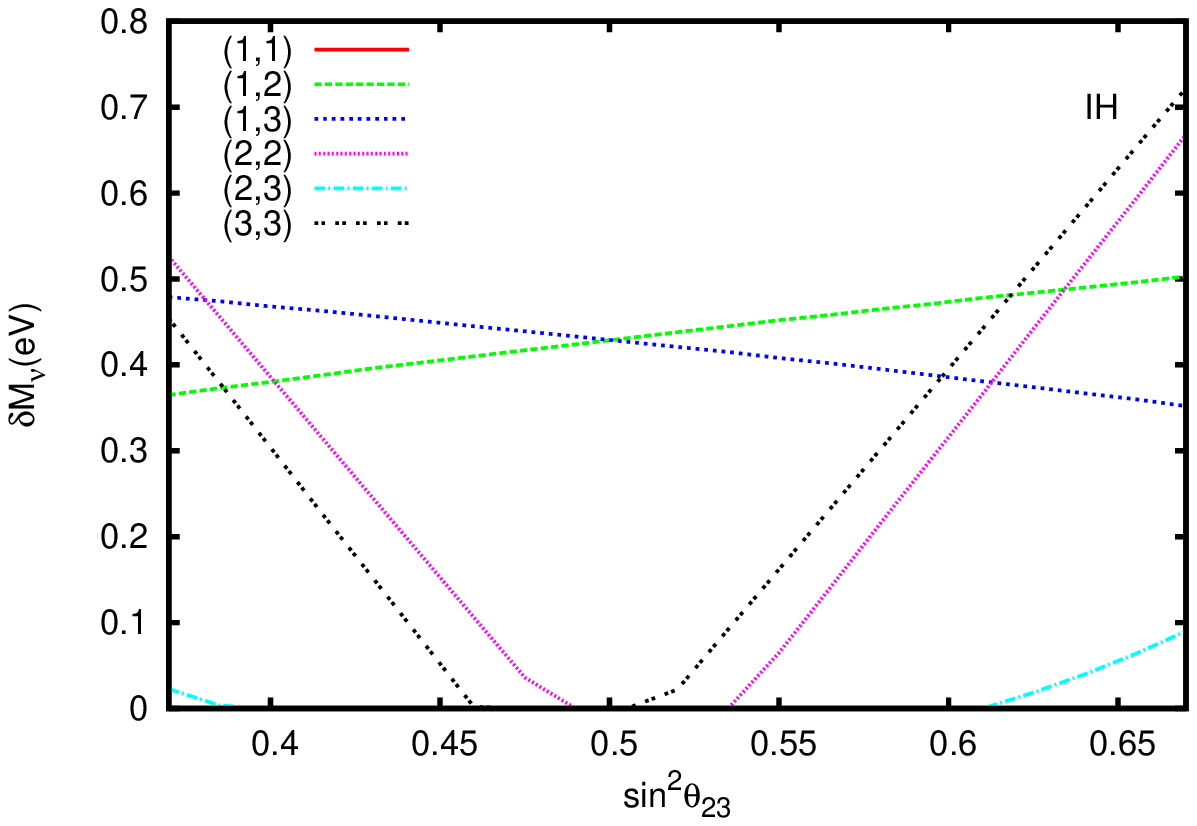}
\includegraphics[width=8cm,height=5cm]{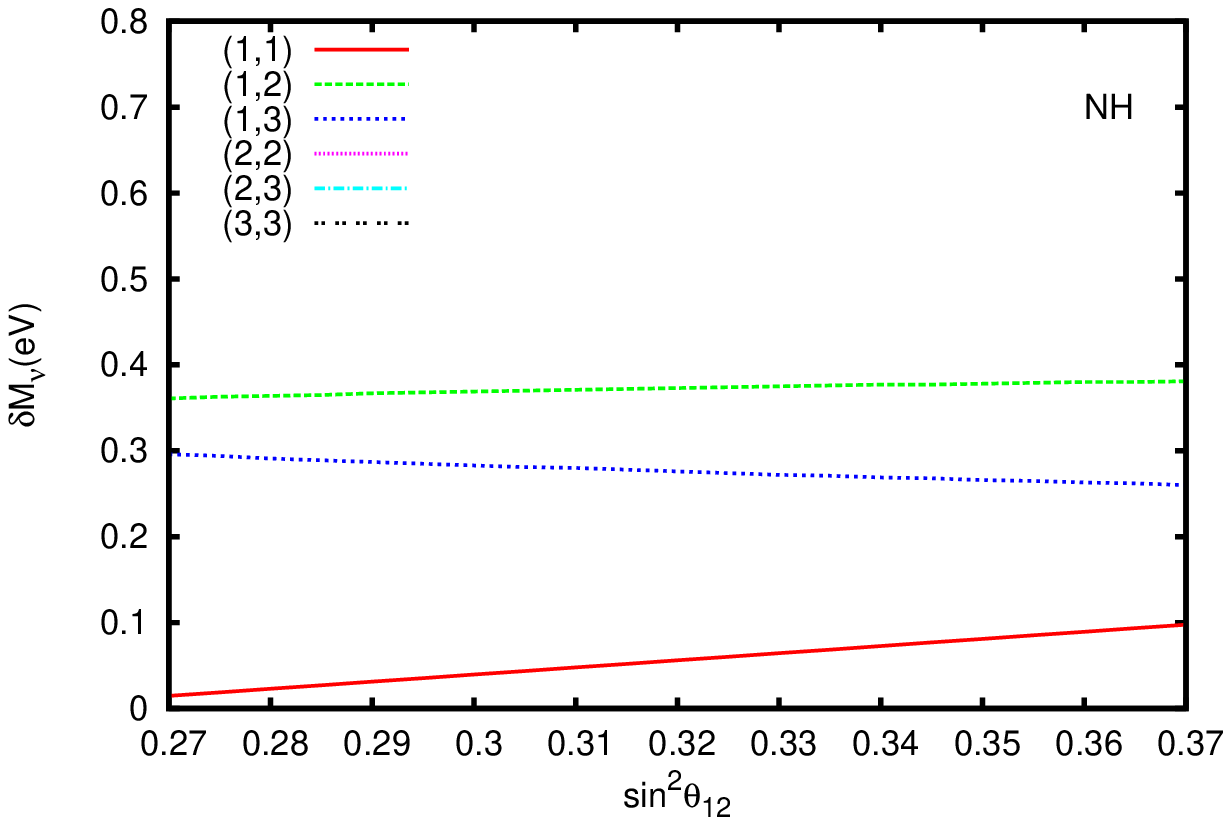}
\includegraphics[width=8cm, height=5cm]{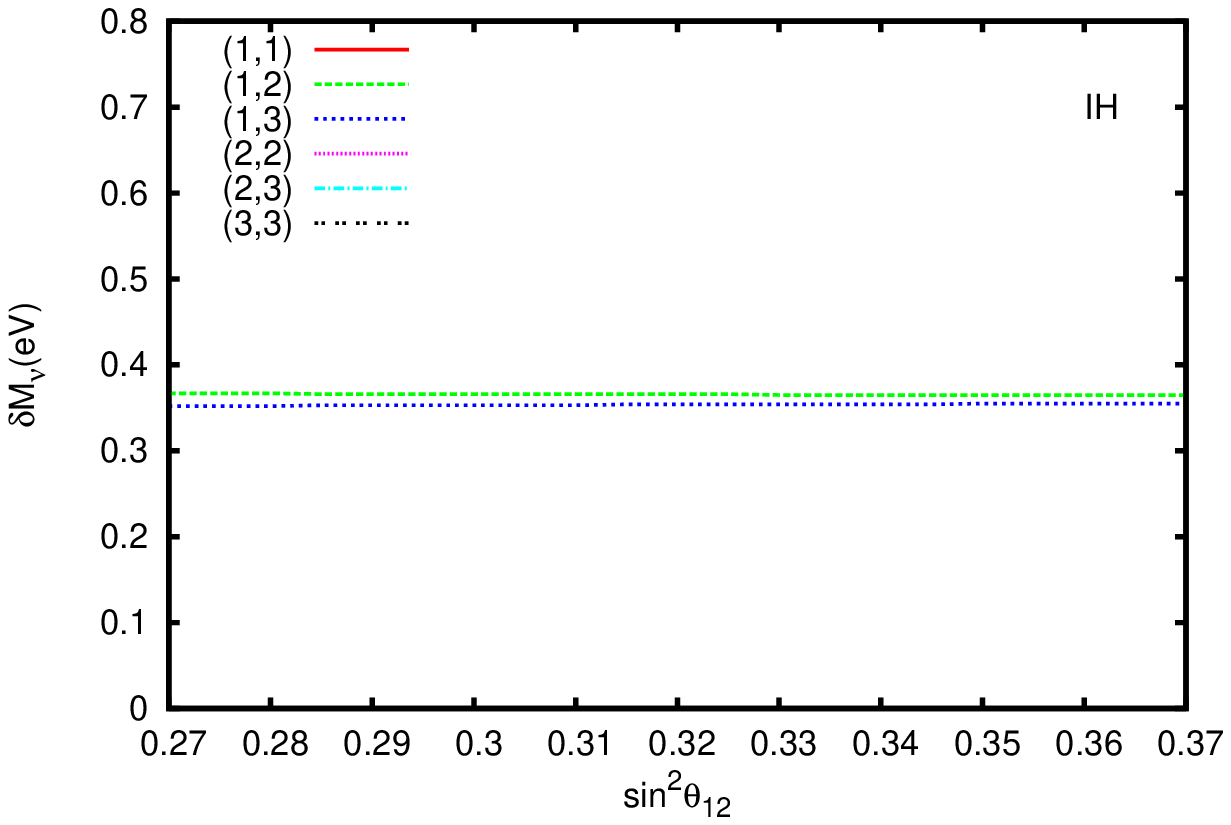}
\end{center}
\caption{$(\delta M_\nu(i,j)\times 10^2)$ as function of mixing angles for $m_0 = 0$ in case of NH (left) and IH (right) 
of neutrino mass matrix.} 
\label{mnuij}
\end{figure}

We now try to explore the $\theta_{13}$, $\theta_{23}$, and $\theta_{12}$ dependency of all the elements of the neutrino 
mass matrix in the flavor basis where the charged lepton is assumed to be diagonal. In order to see the effect of a particular 
oscillation parameter on the value of $\delta M_{\nu}$, we vary that parameter in the $3\sigma$ allowed range and
do a marginalization over the other oscillation parameters. From Eq.~(\ref{delta_mu_nu}) we see that in either case of NH and 
IH the dependency of $\delta M_\nu(1,2)$ and $\delta M_\nu(1,3)$ on $\theta_{12}$ is almost negligible because of the cancellation 
in the second term of each expression. However, both $\delta M_\nu(1,2)$ and $\delta M_\nu(1,3)$ do depend on $\theta_{23}$ and 
$\theta_{13}$ in either case of NH and IH as can be seen from Fig.~\ref{mnuij}.  
The dependency of $\delta M_\nu(1,2)$ and $\delta M_\nu(1,3)$ elements on $\theta_{13}$ and $\theta_{23}$ can be 
understood from the following analytical approximation. In case of NH, $m_0 \equiv m_1 \to 0$. Hence we have $m_3 = 
\sqrt{\Delta m^2_{\rm atm}}$ and $m_2=\sqrt{\Delta m^2_{\odot}}$. As a result from Eq. (\ref{delta_mu_nu}) we get 
\begin{eqnarray} 
|\delta M_\nu(1,2)| &=& \left[ (s_{12}c_{13}(c_{12}c_{23}- s_{12}s_{23}s_{13})-1/3)\sqrt{\Delta m^2_{\odot}} 
+ s_{13}c_{13}s_{23}\sqrt{\Delta m^2_{\rm atm}} \right]\,{\rm Exp}[-m_0/a] \nonumber\\
|\delta M_\nu(1,3)| &=& \left[ (s_{12}c_{13}(-c_{12}s_{23}-s_{12}c_{23}s_{13}) + 1/3) \sqrt{\Delta m^2_{\odot}} 
+ s_{13}c_{13}c_{23}\sqrt{\Delta m^2_{\rm atm}} \right]\,{\rm Exp} [-m_0/a]
\end{eqnarray}
where the exponential factor in $|\delta M_\nu(1,2)|$ and $|\delta M_\nu(1,3)|$ gives the lightest neutrino mass dependency. 
Because of the cancellation, the 1st term in the square bracket is always suppressed in comparison to the second term. Therefore, 
we get: 
\begin{eqnarray}
|\delta M_\nu(1,2)| & \approx &  \sqrt{\Delta m^2_{\rm atm}}  \sin \theta_{13} \cos\theta_{13} \sin \theta_{23}\,{\rm Exp} [-m_0/a] \nonumber\\
|\delta M_\nu(1,3)| &\approx &  \sqrt{\Delta m^2_{\rm atm}} \sin \theta_{13} \cos\theta_{13} \cos \theta_{23}\,{\rm Exp} [-m_0/a]\,. 
\label{12-13-dependency}
\end{eqnarray} 
On the other hand in case of IH, $m_0\equiv m_3 \to 0$. Hence we have $m_2\approx m_1= \sqrt{\Delta m^2_{\rm atm}}$. 
As a result, in the same analogy of Eq. (\ref{12-13-dependency}), we get
\begin{eqnarray} 
|\delta M_\nu(1,2)| & \approx &  \sqrt{\Delta m^2_{\rm atm}}  \sin \theta_{13} \cos\theta_{13} \sin \theta_{23}\,{\rm Exp} [-m_0/a] \nonumber\\
|\delta M_\nu(1,3)| &\approx &  \sqrt{\Delta m^2_{\rm atm}} \sin \theta_{13} \cos\theta_{13} \cos \theta_{23}\,{\rm Exp} [-m_0/a]\,. 
\end{eqnarray}
Thus in either cases the dependency of $\delta M_\nu(1,2)$ and $\delta M_\nu(1,3)$ elements on $\theta_{13}$ and $\theta_{23}$ 
are very similar. This can be checked from Fig.~\ref{mnuij}. It is evident from Fig.~\ref{mnuij} that $\delta M_\nu(2,2)$, 
$\delta M_\nu(2,3)$, and $\delta M_\nu(3,3)$ elements don't depend on $\theta_{12}$. Those elements depend mostly on the value of 
$\theta_{23}$. To understand this feature quantitatively, we write down the matrix elements $\delta M_\nu(2,2)$, 
$\delta M_\nu(2,3)$, and $\delta M_\nu(3,3)$ in the following. For normal ordering of the neutrino mass spectrum, we have 
$m_0\equiv m_1 \to  0$. As a result, in the same analogy of Eq. (\ref{12-13-dependency}), we have:
\begin{eqnarray}
\delta M_\nu(2,2) & \approx & (\cos^2\theta_{13}\,\sin^2\theta_{23} - \frac{1}{2})\,\sqrt{\Delta m^2_{\rm atm}}\,{\rm Exp} [-m_0/a]\,, \nonumber \\
\delta M_\nu(2,3) & \approx & (\sin\theta_{23}\,\cos\theta_{23}\,\cos^2\theta_{13} - \frac{1}{2})\,\sqrt{\Delta m^2_{\rm atm}}\,{\rm Exp} [-m_0/a]\,, \nonumber \\
\delta M_\nu(3,3) & \approx & (\cos^2\theta_{13}\,\cos^2\theta_{23} - \frac{1}{2})\,\sqrt{\Delta m^2_{\rm atm}}\,{\rm Exp} [-m_0/a]
\end{eqnarray}
Clearly, $\delta M_\nu(2,2)$, $\delta M_\nu(2,3)$, and $\delta M_\nu(3,3)$ don't depend on $\theta_{12}$ at all. There is a 
$\theta_{13}$ dependency and is of similar order if $\theta_{23}$ is near to its TBM value. However, once the value of $\theta_{23}$ 
deviates away from its TBM value, all these matrix elements mainly depend on the value of $\theta_{23}$. Moreover, all these 
$\delta M_\nu(i,j)$ increases as we move away from the TBM value. In case of IH, $m_0 \equiv m_3 \to 0$ and a similar pattern
for all the $\delta M_{\nu}(i,j)$ is expected. This can be easily read from the right panel of Fig.~\ref{mnuij}.

\section{Application to Top-down approach}
\label{model}
Now that we know the perturbed matrix from the bottom-up approach, we wish to construct models of neutrino masses 
and mixings in a top-down approach and elucidate the role of lightest neutrino mass. Our main objective here is that 
we will use the predictions of bottom-up approach as guide lines for building neutrino mass matrix whose tree-level 
mixing is governed by a symmetry. We then modify this using some perturbation so that the resulting values are consistent 
with the results that are obtained using the bottom up approach of section.~\ref{tbm}. We begin by proposing a model 
where the neutrino mixing is described by $U_M = U_0\, V$, where $V$ is the perturbed mixing matrix around the tree-level 
value and in general is given by:
\begin{eqnarray}
V=V(1,2)\,V(1,3)\,V(2,3)\,,
\end{eqnarray}
with $V(i,j)$ being given by an orthogonal rotation matrix in the $(i,j)$ plane of neutrino mass matrix. Let us 
write down $V(i,j)$ explicitly such that the perturbed mixing matrix is
\begin{eqnarray}
V = V(1,2)\,V(1,3)\,V(2,3) &=&
\left( \begin{array}{ccc}
1 & \beta  & 0 \\
-\beta & 1 & 0 \\
0 & 0 & 1 \end{array} \right)
\left( \begin{array}{ccc}
1 & 0  & \alpha \\
0 & 1 & 0 \\
-\alpha & 0 & 1 \end{array} \right)
\left( \begin{array}{ccc}
1 & 0  & 0 \\
0 & 1 & \gamma \\
0 & -\gamma & 1 \end{array} \right)
 \nonumber \\
 &&=
\left( \begin{array}{ccc}
1 & \beta - \alpha\gamma  & \alpha + \beta\gamma \\
-\beta & 1 & \gamma -\beta\alpha \\
-\alpha & -\gamma & 1 \end{array} \right),
\end{eqnarray}
where the determinant of $V(i,j)$ matrix is assumed to be unity. Since $U_M$ is the diagonalising matrix of the proposed 
neutrino mass matrix, we get 
\begin{eqnarray}
M_{\nu}^D &=&
U^T_M\,(M_{\nu})_M\,U_M = U^T_M\,((M_{\nu})_{0} + \delta M_{\nu})\,U_M \nonumber \\
&&=
V^T\,U^T_0\,((M_{\nu})_{0} + \delta M_{\nu})\,U_0\,V\,,
\end{eqnarray}
where, $(M_{\nu})_M$ is the mass matrix in a flavor basis where charged leptons are real and diagonal. We now 
define a matrix $X$ such that
\begin{eqnarray}
X &\equiv & V\,M_{\nu}^D\,V^T =
U^T_{0}\,((M_{\nu})_{0} + \delta M_{\nu})\,U_{0}\,,
\label{topdown_bottomup}
\end{eqnarray}
where the right-hand-side is determined through the bottom-up approach for a specific value of $m_0$, where as the left-hand-side 
involves six parameters, namely $m_0$, $\Delta m_{\rm atm}^2$, $\Delta m_{\odot}^2$, $\alpha,\beta$ and $\gamma$. Note that for 
$m_0 < 10^{-3} {\rm eV}$ we can neglect $m_0$ dependency as we discussed earlier in Section.~\ref{tbm}.

\subsection{Application to Tri-bi-maximal mixing}
Let us identify the tree-level mixing matrix as a tri-bi-maximal one, predicted by certain symmetry, then $U_0=U_{\rm TBM}$. 
As a result from Eq. (\ref{topdown_bottomup}) we get 
\begin{eqnarray}
\label{Xex}
X &\equiv & V\,M_{\nu}^D\,V^T =
U^T_{\rm TBM}\,((M_{\nu})_{\rm TBM} + \delta M_{\nu})\,U_{\rm TBM}\,.
\end{eqnarray}
The matrix elements $X(i,j)$ can be expressed in terms of $m_1,\,m_2,\,m_3,\,\alpha,\,\beta,\,\gamma$ as 
\begin{eqnarray}
&&X(1,1) = m_1 + m_2\,\beta^2 + m_3\,\alpha^2 \,, \nonumber \\
&&X(1,2) = (m_2 - m_1)\,\beta + (m_3 - m_2)\alpha\,\gamma \,, \nonumber \\
&&X(1,3) = (m_3 - m_1)\,\alpha\, + (m_3 - m_2)\beta\gamma \,, \nonumber \\
&&X(2,2) = m_1\,\beta^2 + m_2 + m_3\,\gamma^2 \,, \nonumber \\
&&X(2,3) = -(m_3 - m_1)\,\alpha\,\beta\, + (m_3 -m_2)\gamma \,, \nonumber \\
&&X(3,3) = m_1\,\alpha^2 + m_2\gamma^2 + m_3\,,
\label{Xth}
\end{eqnarray}
where $m_2 = \sqrt{m_1^2 + \Delta m_{\odot}^2}$ and $m_3 = \sqrt{m_1^2 + \Delta m_{\rm atm}^2}$, respectively. Here we assume normal ordering of the
neutrino masses. 
We have ignored higher order terms in $\alpha,\,\beta,\,{\rm and}\,\gamma$. We wish to determine the range of $\alpha,\,\beta,\,\gamma$
such that $X(i,j)$ computed from Eq.~(\ref{Xth}) is compatible within the $3\sigma$ ranges of $X(i,j)$ derived from the bottom-up approach, i.e, from the
right hand side of Eq.~(\ref{Xex}), for $m_0 \equiv m_1 = 0$ (i.e., $m_0 < 10^{-3}$ eV) as well as for $m_0 \equiv m_1 \ne 0$ (i.e. $m_0 >10^{-3}$ eV). 

\subsubsection{Case-I: $m_0\equiv m_1 \to 0$}
\begin{figure}[htbp]
\begin{center}
\includegraphics[width=8cm,height=5cm]{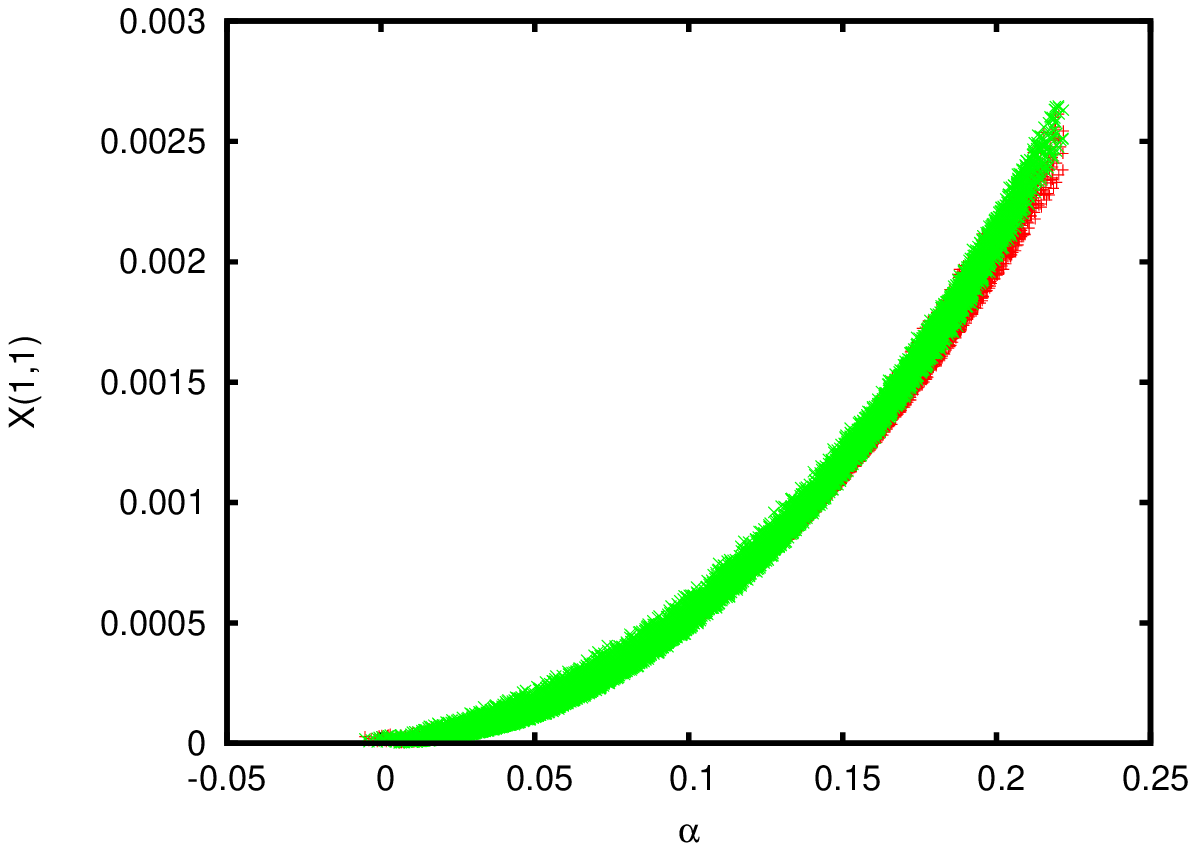}
\includegraphics[width=8cm,height=5cm]{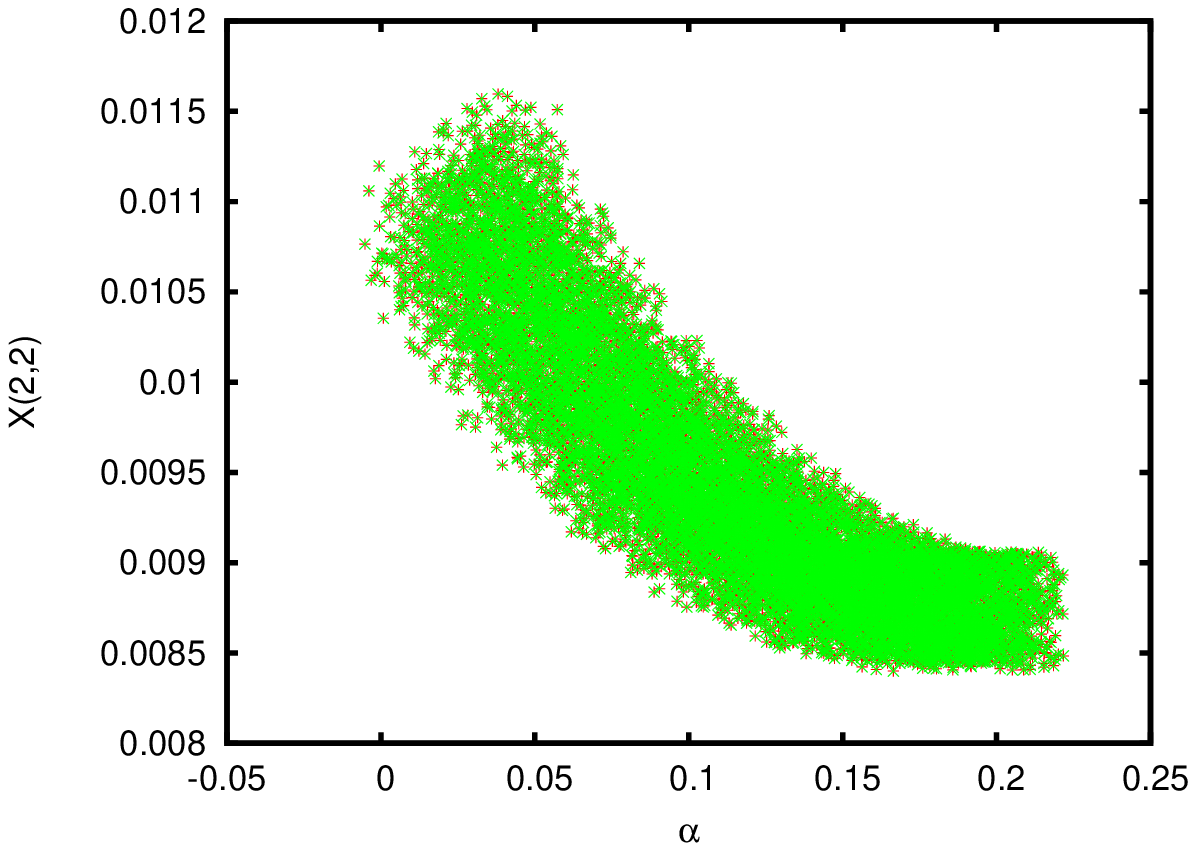}
\includegraphics[width=8cm,height=5cm]{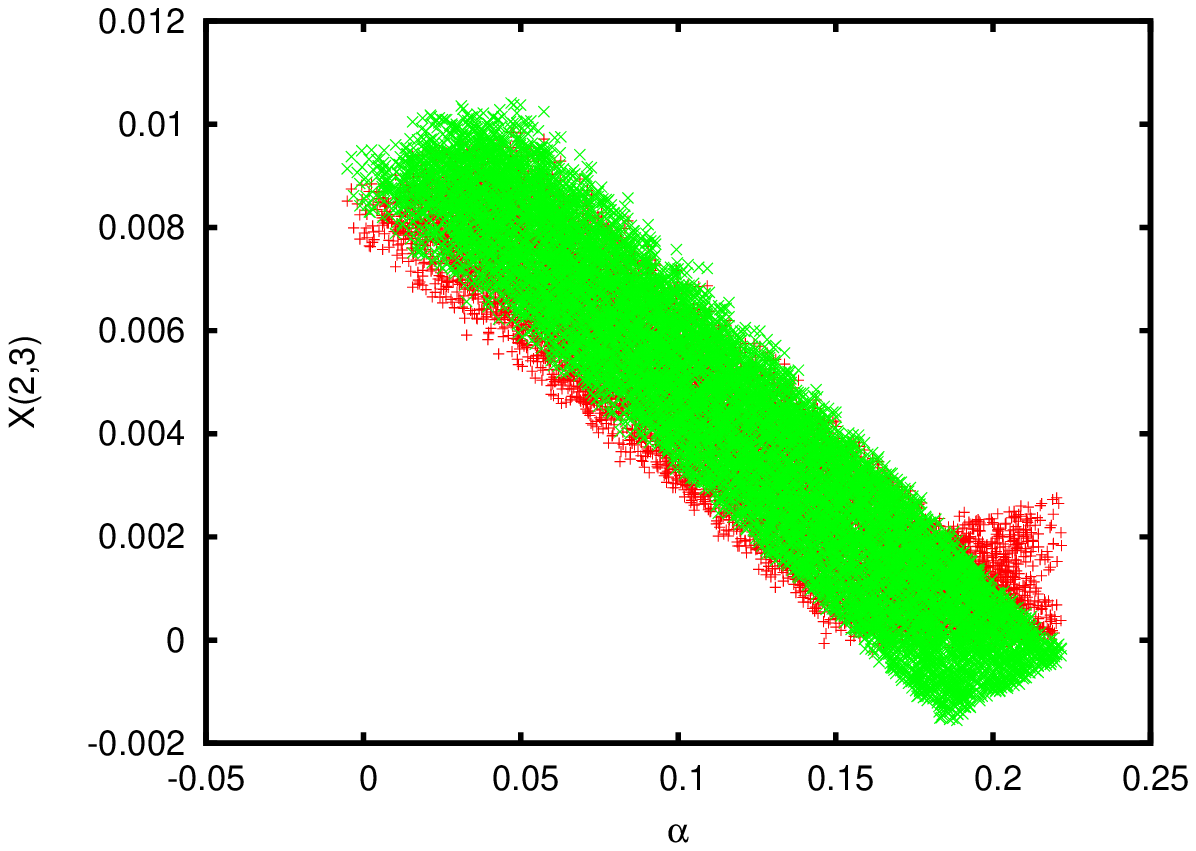}
\includegraphics[width=8cm,height=5cm]{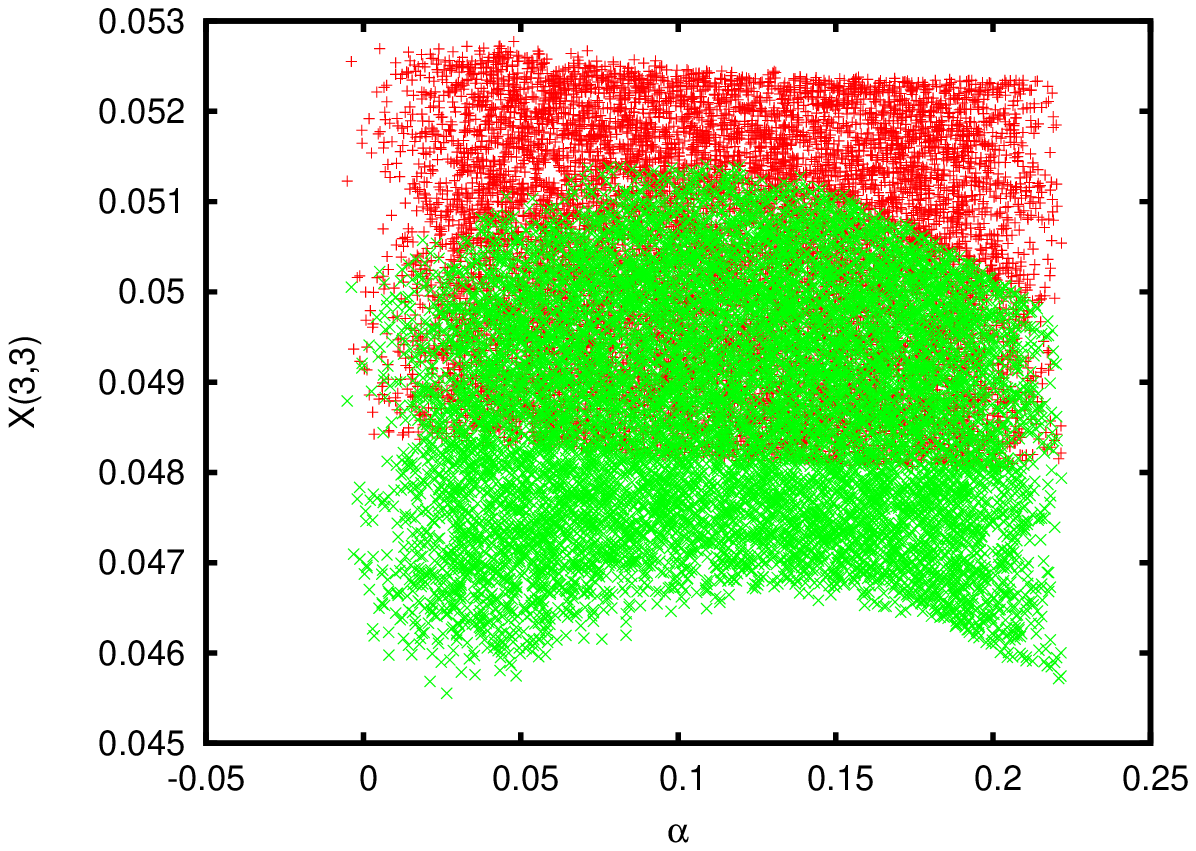}
\end{center}
\caption{Range of $X(i,j)$ as a function of $\alpha$ for $\alpha,\,\beta,\,\gamma \ne 0$.
The theory ranges are denoted by red whereas the experimental ranges from the bottom up approach are denoted by
green.}
\label{Xabg}
\end{figure}

In Table.~\ref{X},
we report all the elements of the matrix $X$ obtained using the bottom up approach for the TBM 
mixing scenario.
\begin{table}[ht]
\centering                          
\begin{tabular}{|c|c|c|c|c|c|c|}           
\hline\hline                        
TBM & X(1,1)/{\rm eV} & X(1,2)/{\rm eV} & X(1,3)/{\rm eV} & X(2,2)/{\rm eV} & X(2,3)/{\rm eV} & X(3,3){\rm eV} \\ [0.5ex]   
\hline                              
Central values & $0.0203$ & $0.0487$ & $0.315$ & $1.01$ & $0.747$ & $4.89$ \\               
$3\sigma$ range & $[0.87\times 10^{-4}, 0.269]$ & $[-0.0937, 0.118]$ & $[0.0141, 1.15]$ & $[0.839, 1.17]$ & $[-0.153, 1.05]$ & $[4.57, 5.14]$ \\ [1ex]         
\hline                              
\end{tabular}
\caption{Values of $X (i,j) \times 10^2$ for TBM mixing from the bottom up approach for $m_0 = 0$.}   
\label{X}          
\end{table}
We need to find the range of $\alpha$, $\beta$, and $\gamma$ so that $X(i,j)$ values computed from Eq.~(\ref{Xth}) are consistent with
the values reported in Table.~\ref{X}. We get the range of $\alpha$, $\beta$, and $\gamma$ by fitting $X(1,2)$, $X(1,3)$ and $X(2,2)$
in Eq.~\ref{Xth} with the corresponding $X(1,2)$, $X(1,3)$, and $X(2,2)$ ranges from Table.~\ref{X}. Once the range of $\alpha$, 
$\beta$, and $\gamma$ are known, we then plug those in Eq.~\ref{Xth} to get the range of $X(1,1)$, $X(2,3)$, and $X(3,3)$ and are shown in 
Fig.~\ref{Xabg}. 
Since $\alpha$, $\beta$ and $\gamma$ are non-zero, it is obvious that all the six $X(i,j)$ elements can be fitted within their experimental 
ranges. However, if more parameters are set to zero then it is not clear if all the six $X(i,j)$ elements can be fitted to their experimental 
values.      

In order to realize the dependence of Eq.~(\ref{Xth}) on number of parameters, we analyze six different possibilities: Firstly, we set one of the three 
parameters to zero while keeping other two non zero (3 ways) and secondly, we set two of the three parameters to zero while keeping the third one 
non-zero (3 ways). 

\begin{itemize}
\item First we set $\gamma=0$ and find the range of $\alpha$ and $\beta$ by comparing $X(1,2)$ and $X(1,3)$ with their experimental values given in 
Table.~\ref{X}. Then we check the consistency of remaining four elements and shown in Fig.~\ref{Xab}. It is clear that, although, the resulting 
range of $X(i,j)$s are compatible within their $3\sigma$ ranges, we do not get a very good fit for $X(2,3)$ because $\gamma$ is a perturbation 
in the $(2,3)$ plane. This in turn can be understood from Eq. (\ref{Xth}) by setting $m_1=0$ and $\gamma=0$. Since $X(2,3) \propto m_3 \alpha 
\beta$ and $\alpha$ is required to be small from fitting with $X(1,3)$, it is clear that theoretical values of $X(2,3)$ has minimal overlap with its 
experimental range of values.
\begin{figure}[htbp]
\begin{center}
\includegraphics[width=8cm, height=5cm]{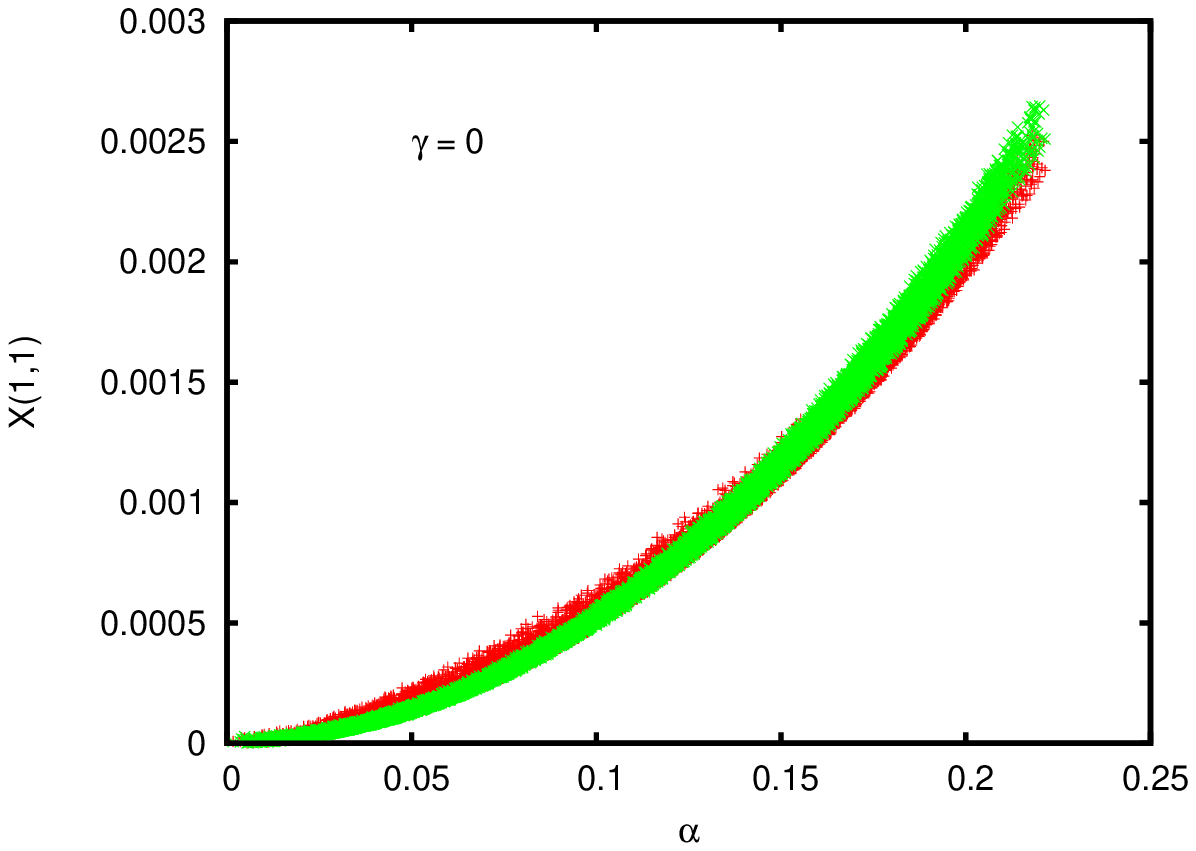}
\includegraphics[width=8cm, height=5cm]{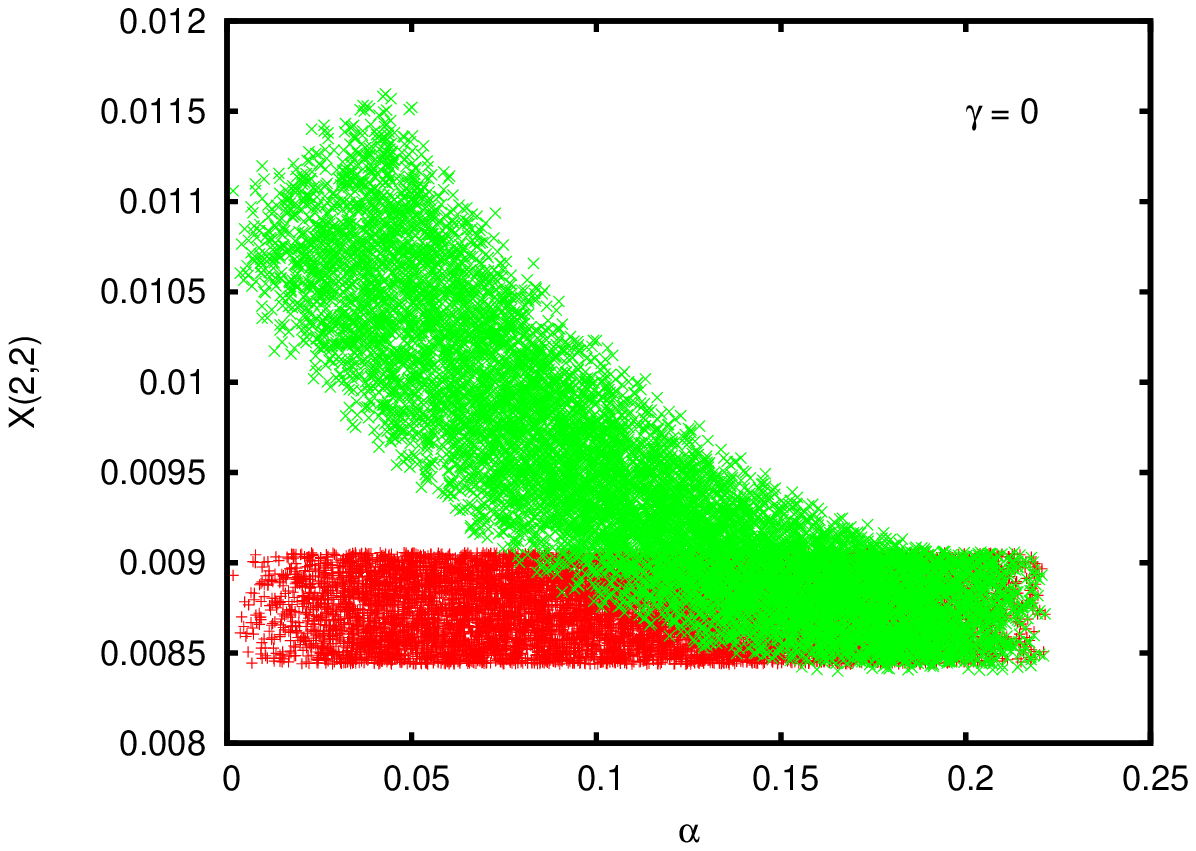}
\includegraphics[width=8cm, height=5cm]{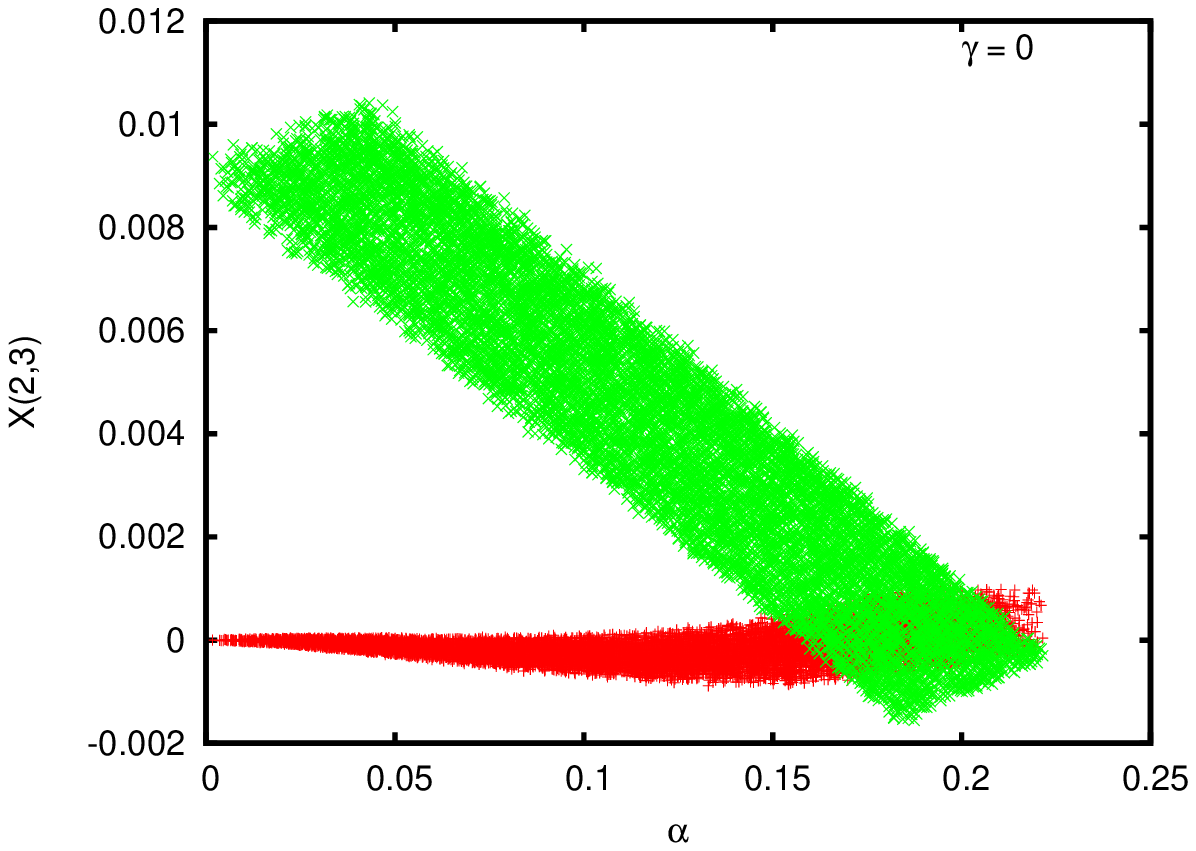}
\includegraphics[width=8cm, height=5cm]{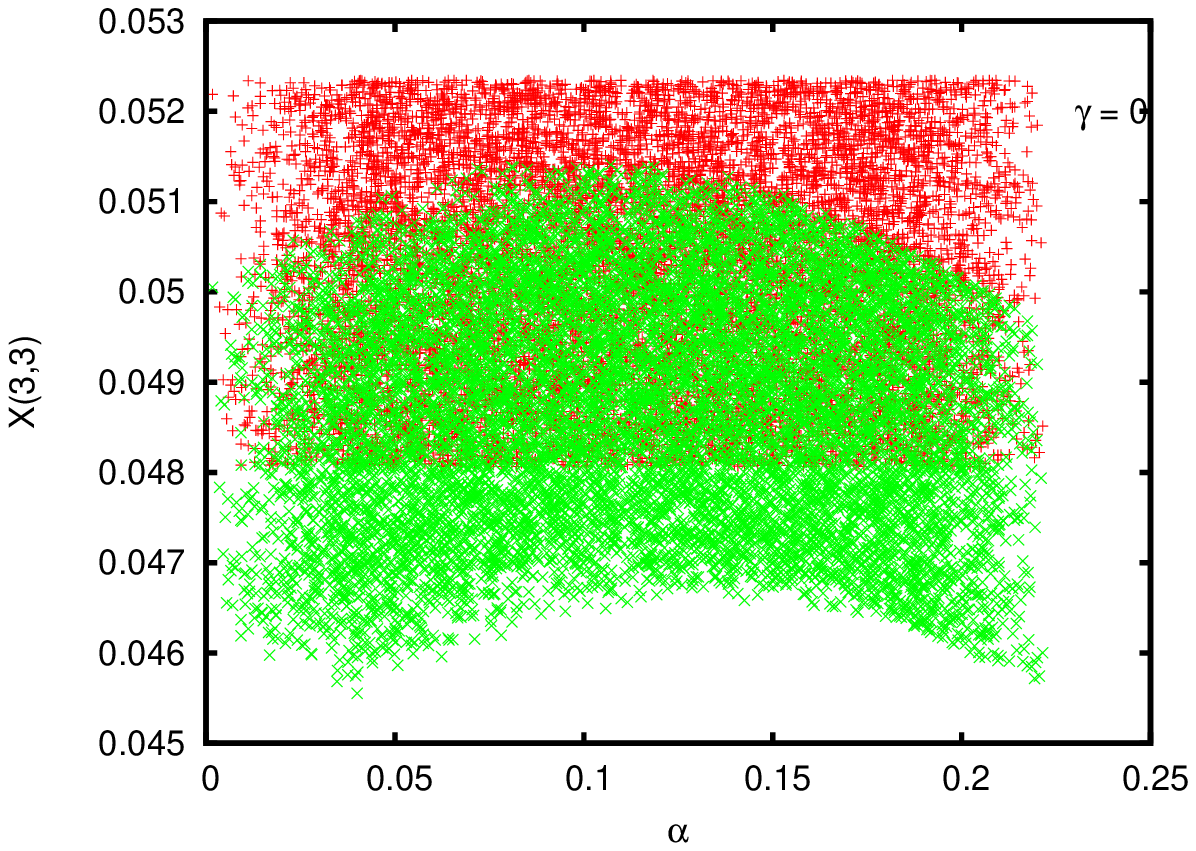}
\end{center}
\caption{Range of $X(i,j)$ as a function of $\alpha$ with $\gamma = 0$.
The theory ranges are denoted by red whereas the experimental ranges from the bottom up approach are denoted by
green.}
\label{Xab}
\end{figure}

\item We perform the same exercise by setting $\beta=0$. We fix the range of $\alpha$ and $\gamma$ from $X(1,3)$ and $X(2,2)$. 
Once the range of $\alpha$ and $\gamma$ are obtained, we then check the consistency of remaining four elements $X(1,1)$, $X(1,2)$, $X(2,3)$, 
$X(3,3)$.  This is shown in Fig.~\ref{Xag}. Since $\beta$ gives the perturbation in $(1,2)$ plane and the 
perturbations have minimal dependencies on $\beta$, we expect a better fit for all the elements. In fact this can be 
quickly checked from Eq.~(\ref{Xth}) by setting $m_1=0$ and $\beta=0$.
\begin{figure}[htbp]
\begin{center}
\includegraphics[width=8cm, height=5cm]{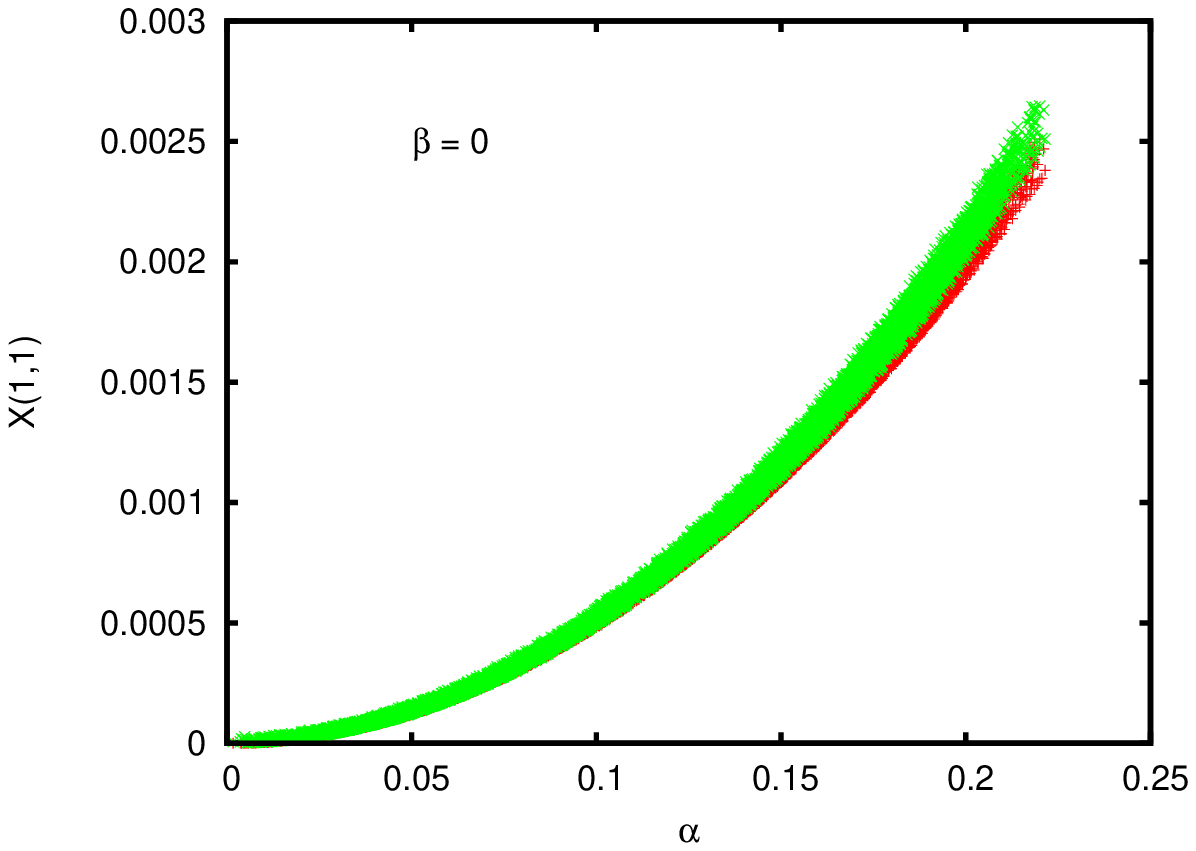}
\includegraphics[width=8cm, height=5cm]{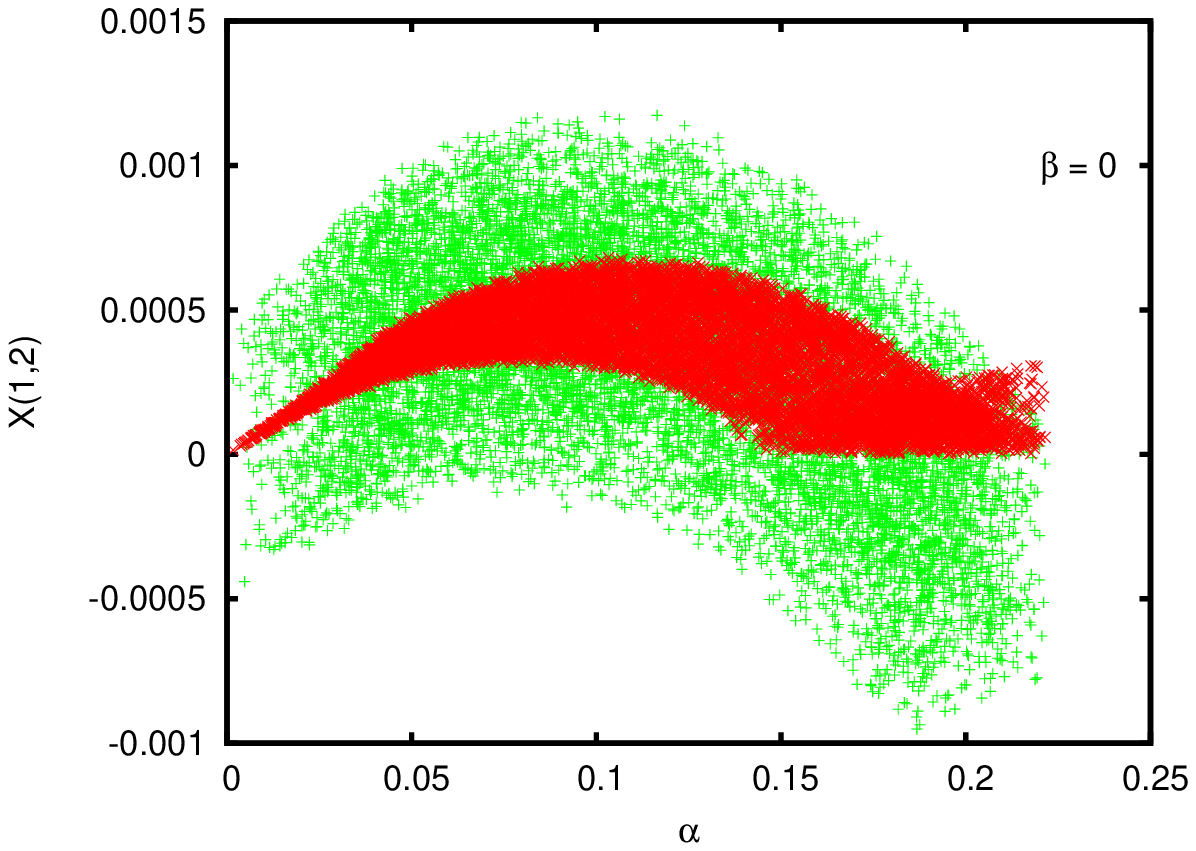}
\includegraphics[width=8cm, height=5cm]{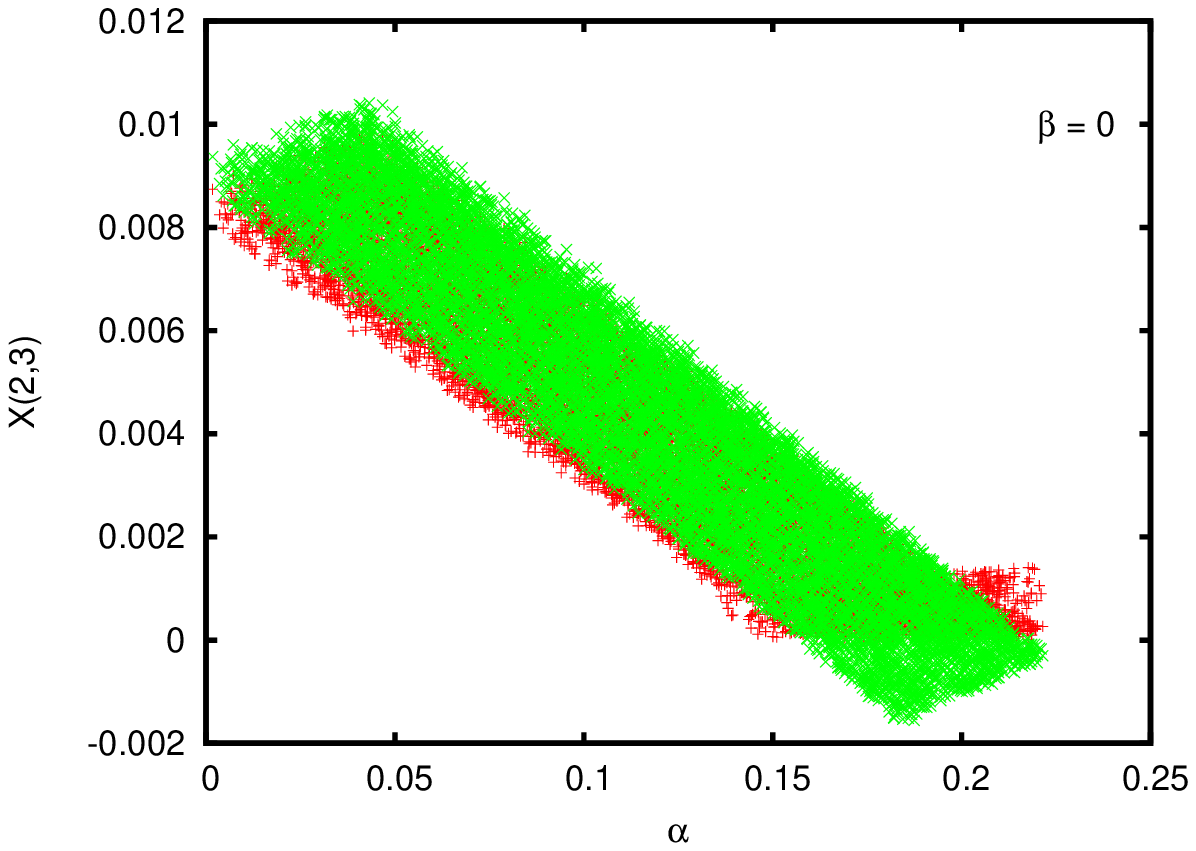}
\includegraphics[width=8cm, height=5cm]{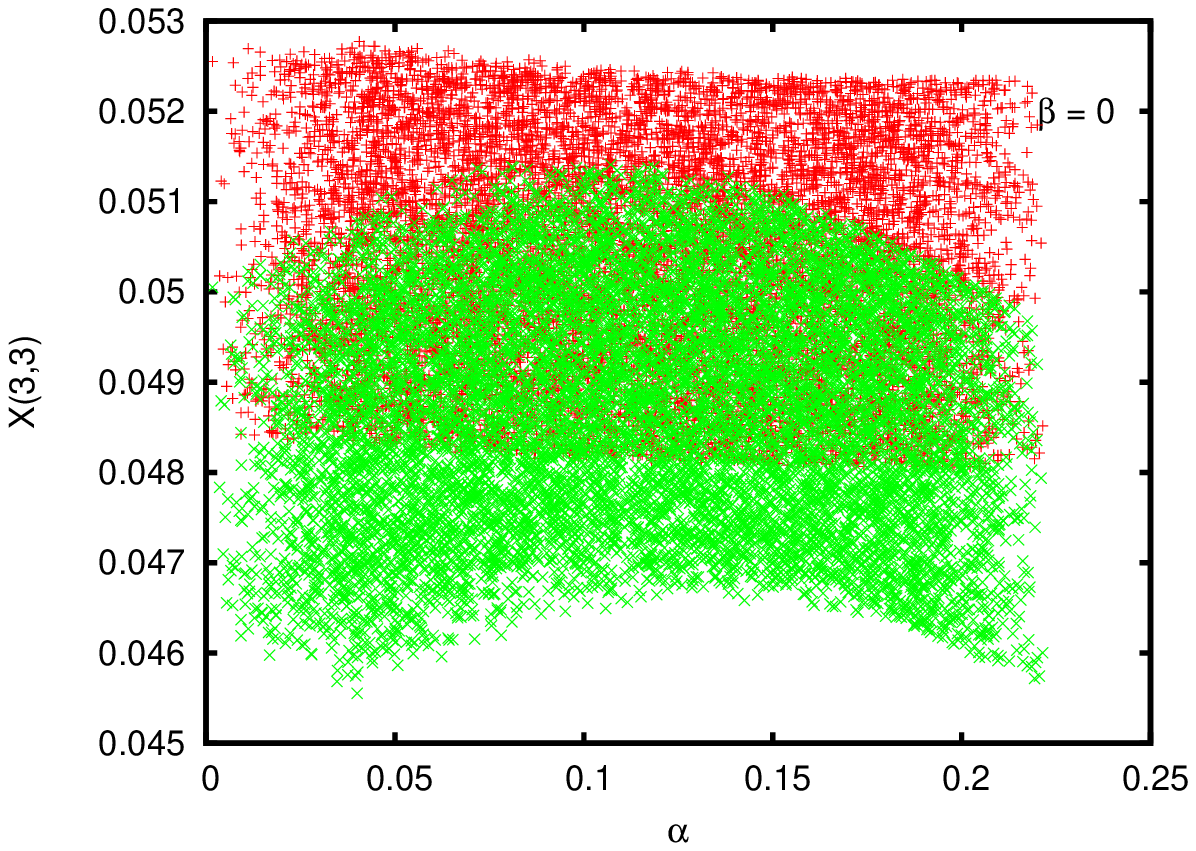}
\end{center}
\caption{Range of $X(i,j)$ as a function of $\alpha$ with $\beta = 0$.
The theory ranges are denoted by red whereas the experimental ranges from the bottom up approach are denoted by
green.}
\label{Xag}
\end{figure}

\item Next we set $\alpha = 0$ and fix the range of $\beta$ and $\gamma$ from $X(1,2)$ and $X(2,2)$. Then we 
check the consistency of remaining four elements $X(1,1)$, $X(1,3)$, $X(2,3)$, $X(3,3)$ and are shown in Fig.~\ref{Xbg}. 
Since $\alpha$ is the perturbation in $(1,3)$ plane and it is set to zero, it is obvious that the theoretical values $X(1,3)$ 
has a minimal overlap with its $3\sigma$ range of experimental values. Note that $X(1,3)$ is not exactly zero even if $\alpha=0$. 
This in fact can be understood from Eq.~(\ref{Xth}) by setting $m_1=0$ and $\alpha=0$. In this limit $X(1,3)=(m_3-m_2)\beta\gamma$ which is not 
zero. Thus we can get a required $\theta_{13}$ by doing perturbation in $(1,2)$ and $(2,3)$ plane. 
\begin{figure}[htbp]
\begin{center}
\includegraphics[width=8cm, height=5cm]{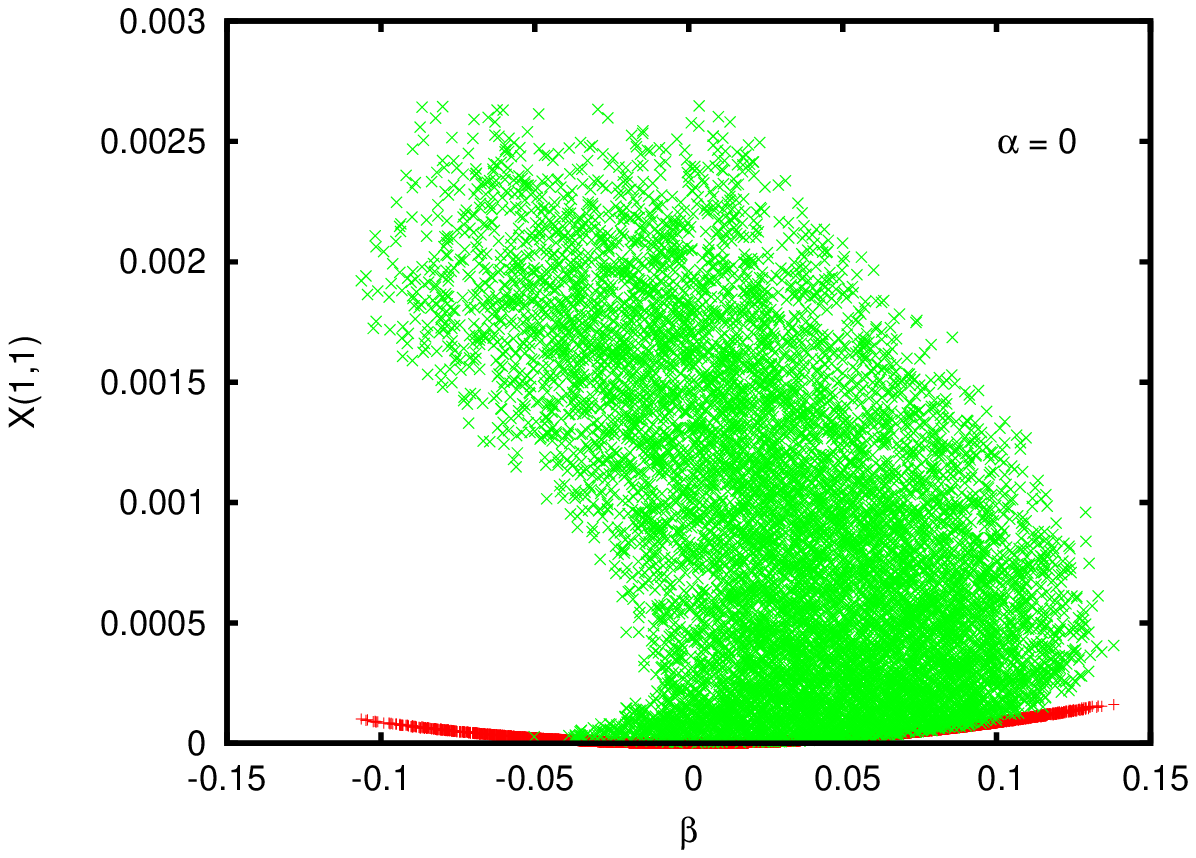}
\includegraphics[width=8cm, height=5cm]{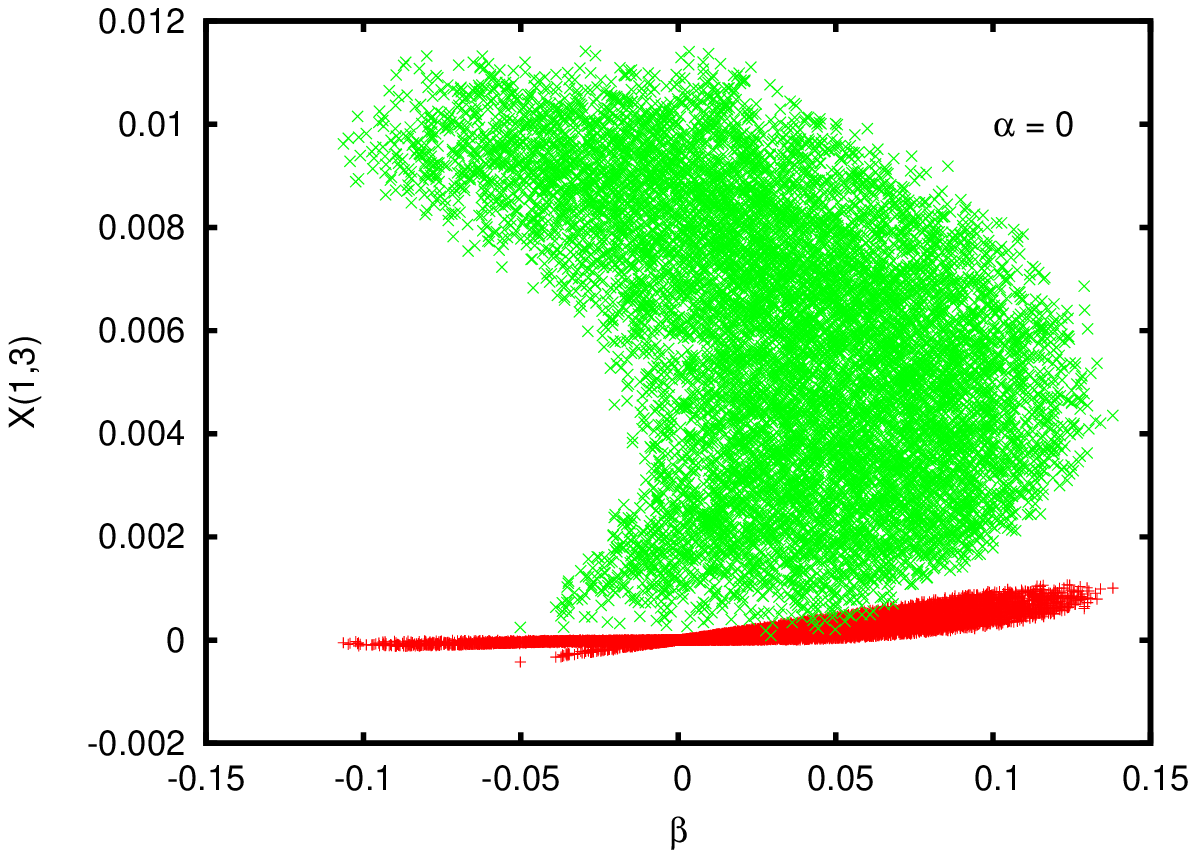}
\includegraphics[width=8cm, height=5cm]{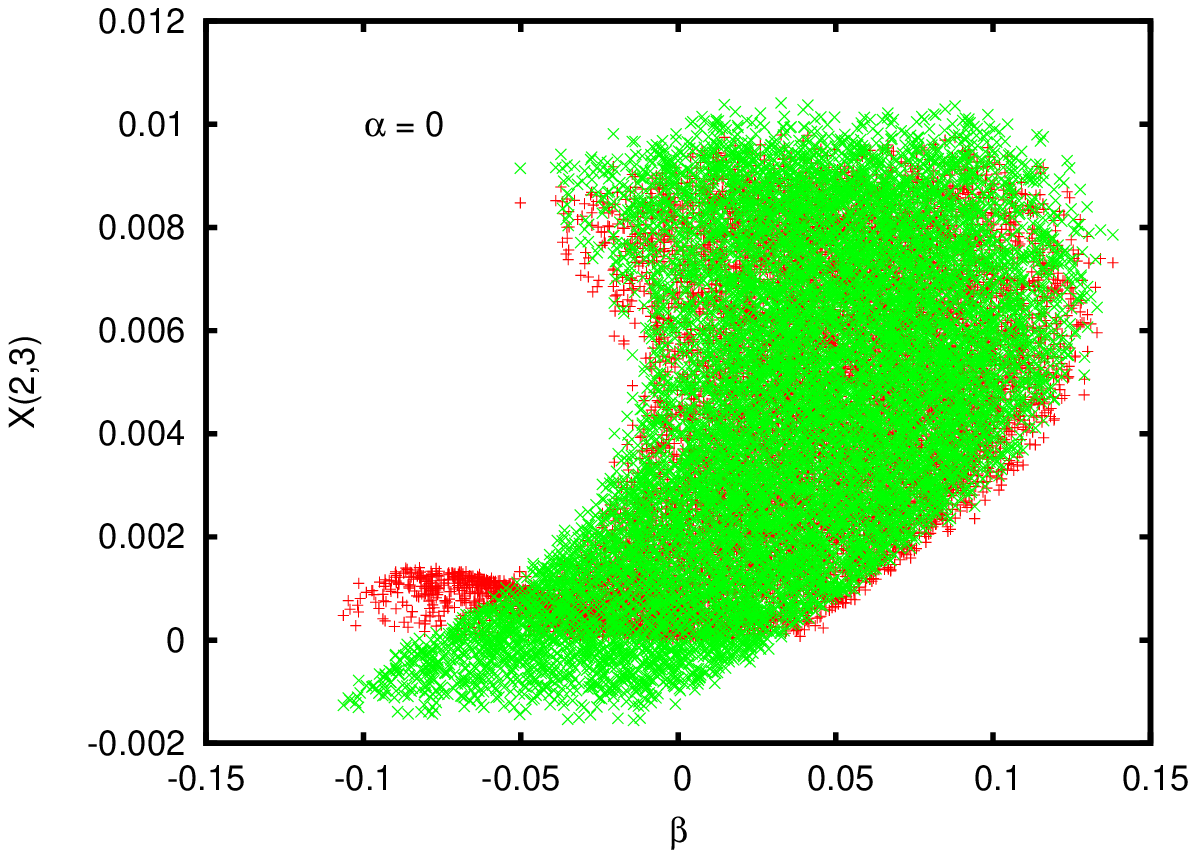}
\includegraphics[width=8cm, height=5cm]{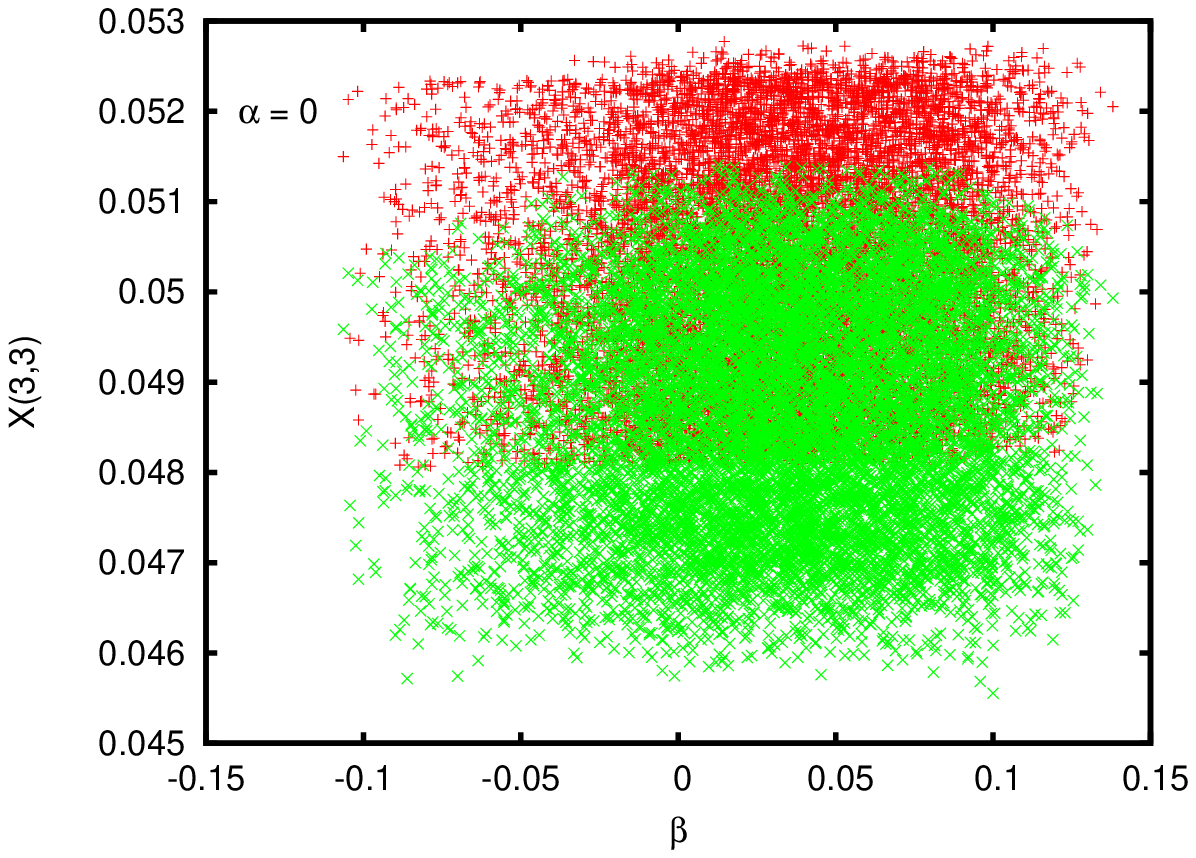}
\end{center}
\caption{Range of $X(i,j)$ as a function of $\beta$ for $\alpha = 0$.
The theory ranges are denoted by red whereas the experimental ranges from the bottom up approach are denoted by
green.}
\label{Xbg}
\end{figure}

\end{itemize}


Since we can set $\alpha$, $\beta$, and $\gamma$ to be zero individually, the order in which we do the rotations to construct the perturbed matrix 
$V$ is not very important. We further note that if $V$ is product of at least two matrices (that is $V$ is parameterized by at least two variables) 
then the parameters in $V$-matrix can not be considered as true-perturbation around their tree-level values. Therefore, the new parameters $\alpha, 
\beta, \gamma$... are arbitrary.

\begin{figure}[htbp]
\begin{center}
\includegraphics[width=5cm, height=4cm]{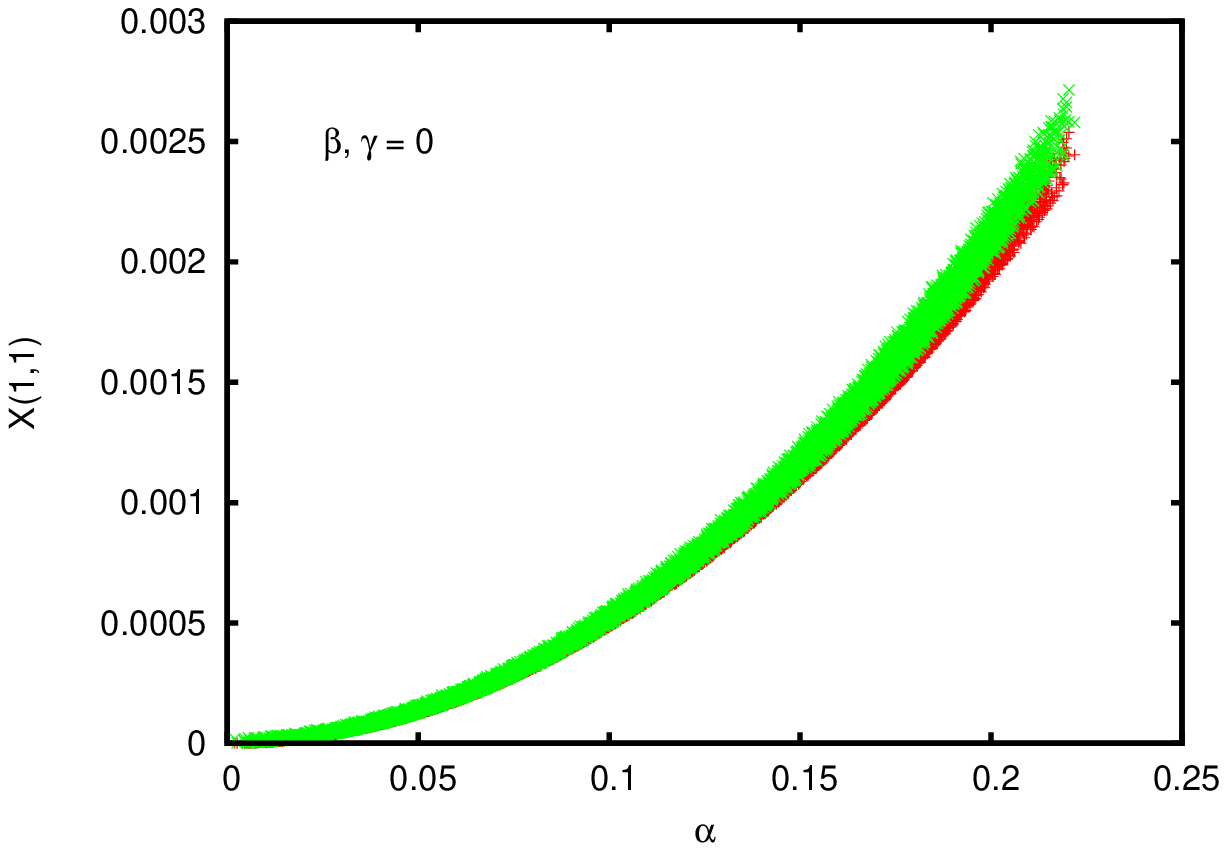}
\includegraphics[width=5cm, height=4cm]{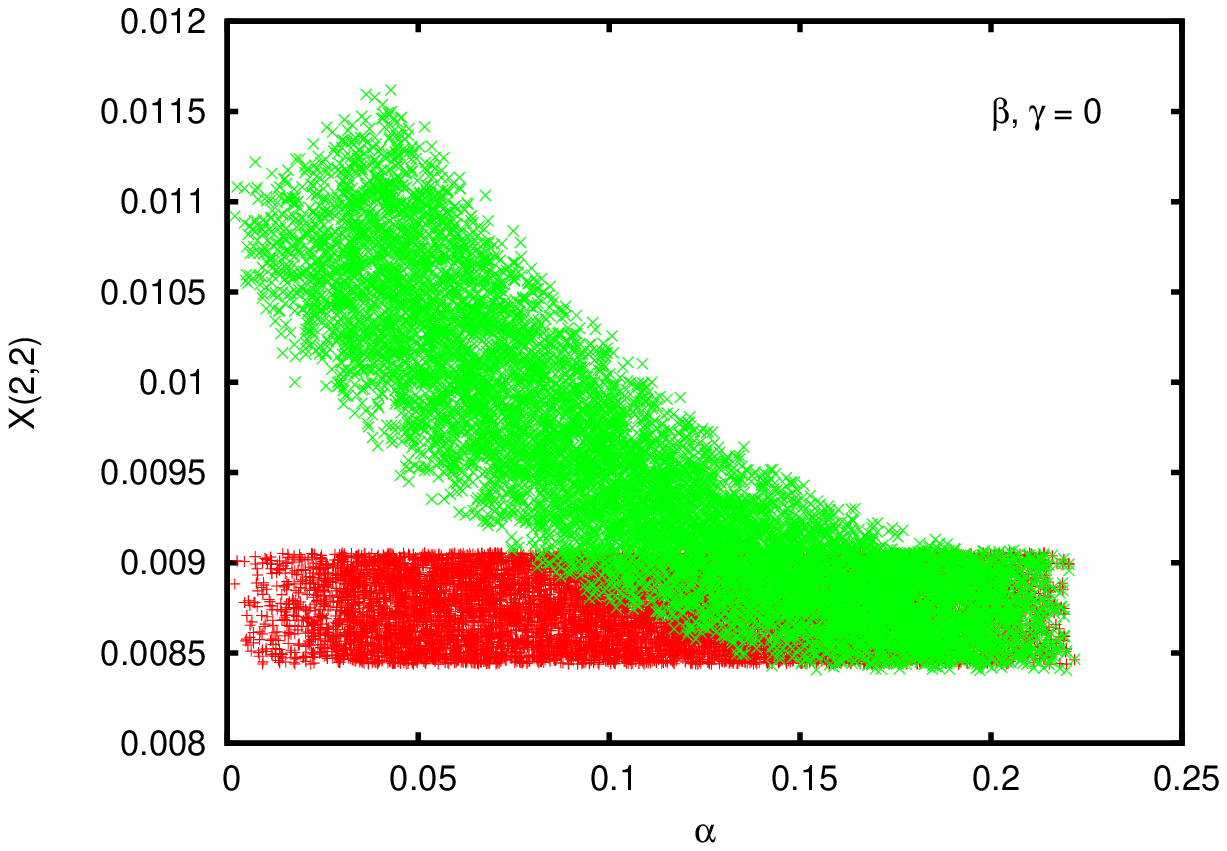}
\includegraphics[width=5cm, height=4cm]{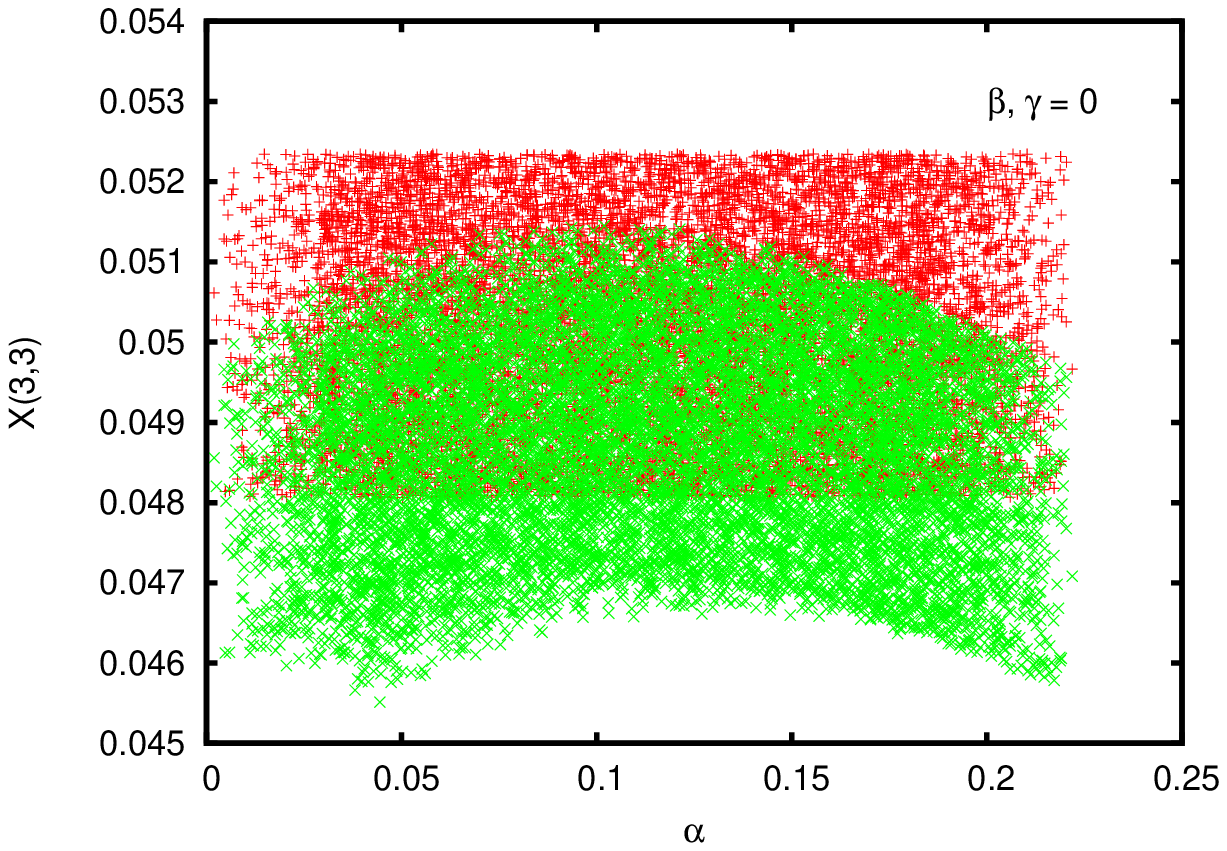}
\end{center}
\caption{Range of $X(i,j)$ as a function of $\alpha $ for $(\beta,\,\gamma = 0)$.
The theory ranges are denoted by red whereas the experimental ranges from the bottom up approach are denoted by
green.}
\label{xalpha}
\end{figure}

Now let us concentrate on the second case where we put two model parameters to zero while keeping third one to be non-zero. It is clear from 
Eq.~(\ref{Xth}) that setting either $(\alpha,\,\beta = 0)$ or $(\alpha,\,\gamma = 0)$ will give $X(1,3) = 0$ which is not compatible with its 
$3\sigma$ range of experimental values reported in Table.~\ref{X}. However, a non zero $\alpha$ with $(\beta,\,\gamma = 0)$ is still a solution 
and is compatible with the experimental data set. In this case, we fix the range of $\alpha$ from $X(1,3)$ and check the consistency of remaining 
five elements. In the limit $m_1=0$ and $\beta=\gamma=0$, from Eq.~(\ref{Xth}) we see that $X(1,2)=X(2,3)=0$ and are compatible with their $3 \sigma$ 
range of experimental values~\footnote{Such a case has been obtained using an $A_4$ symmetry in a type-II seesaw framework~\cite{tbm}.}. The $\alpha$ 
dependency of remaining elements $X(1,1)$, $X(2,2)$, and $X(3,3)$ are shown in Fig.~\ref{xalpha}.

\subsubsection{Case-II: $m_0\equiv m_1 > 10^{-3} {\rm eV}$}
\begin{figure}[htbp]
\begin{center}
\includegraphics[width=5cm, height=4cm]{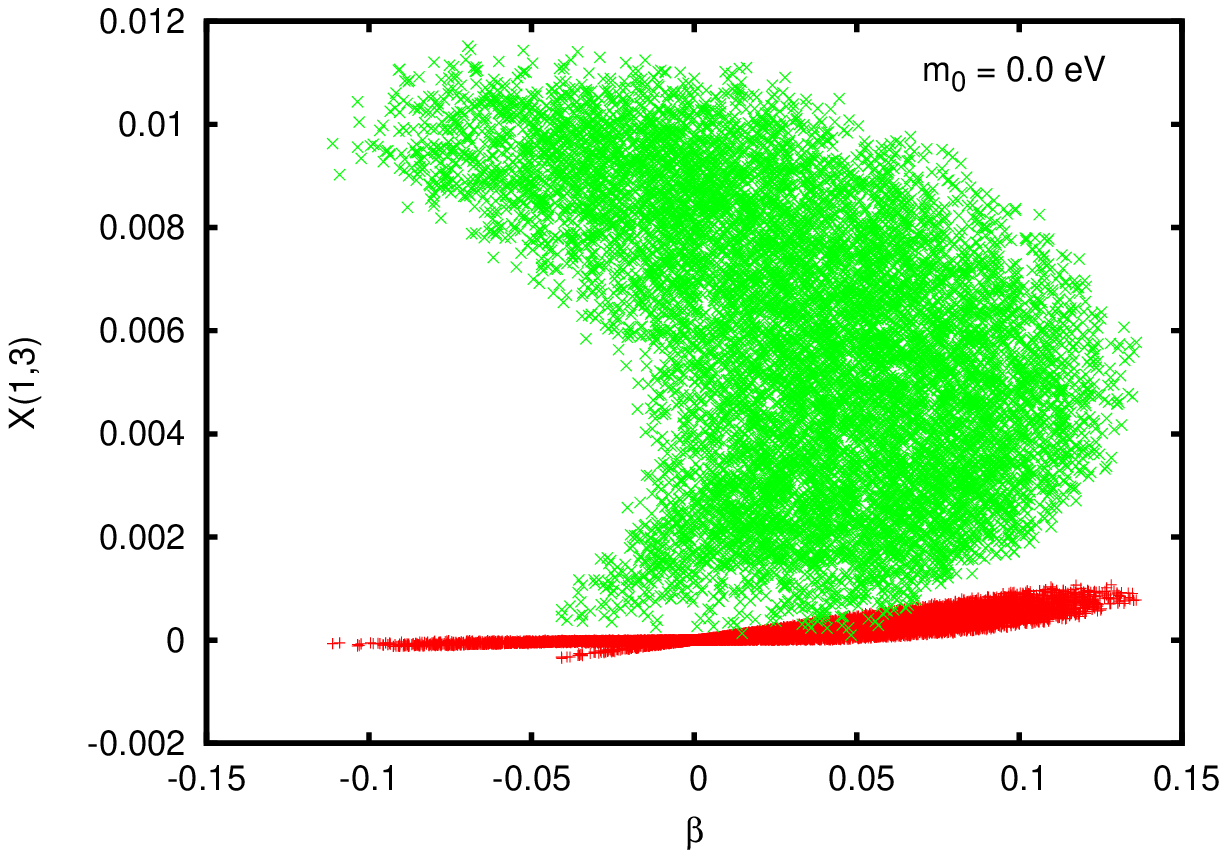}
\includegraphics[width=5cm, height=4cm]{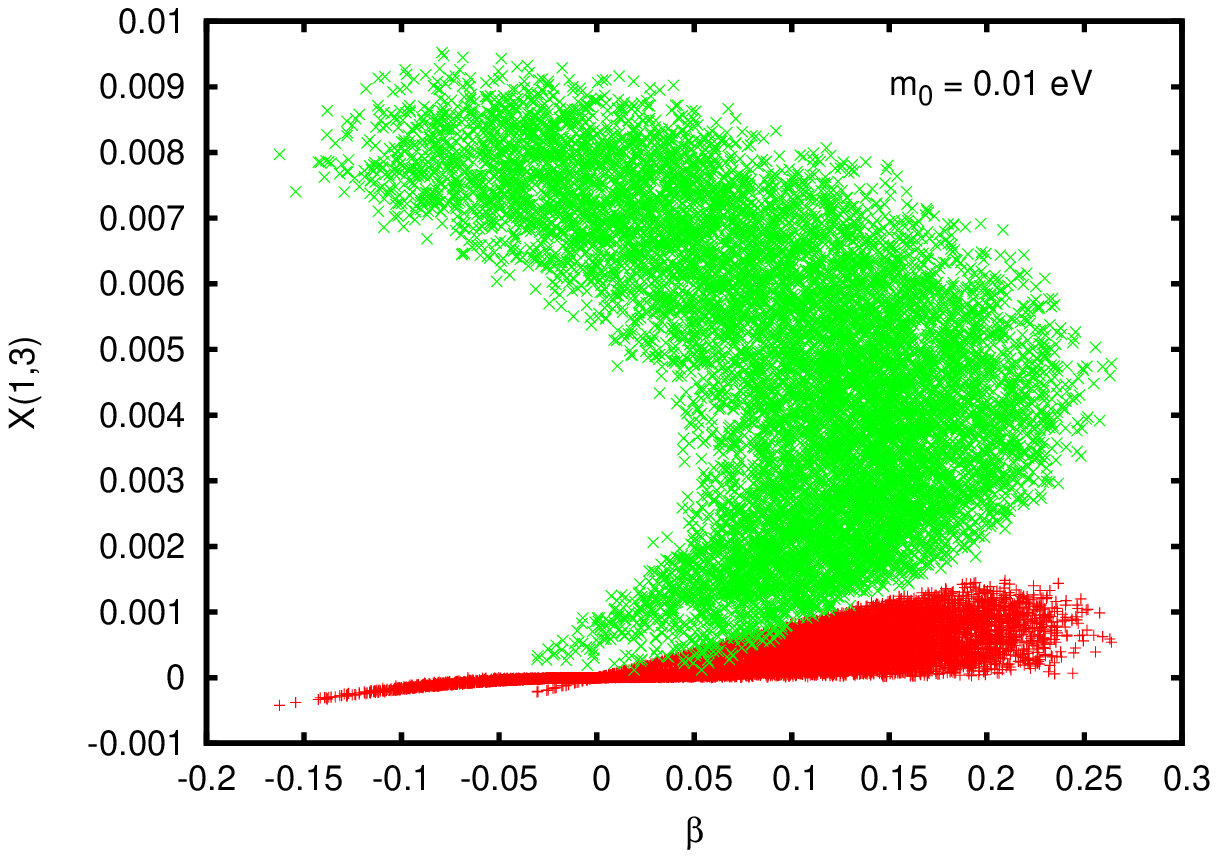}
\includegraphics[width=5cm, height=4cm]{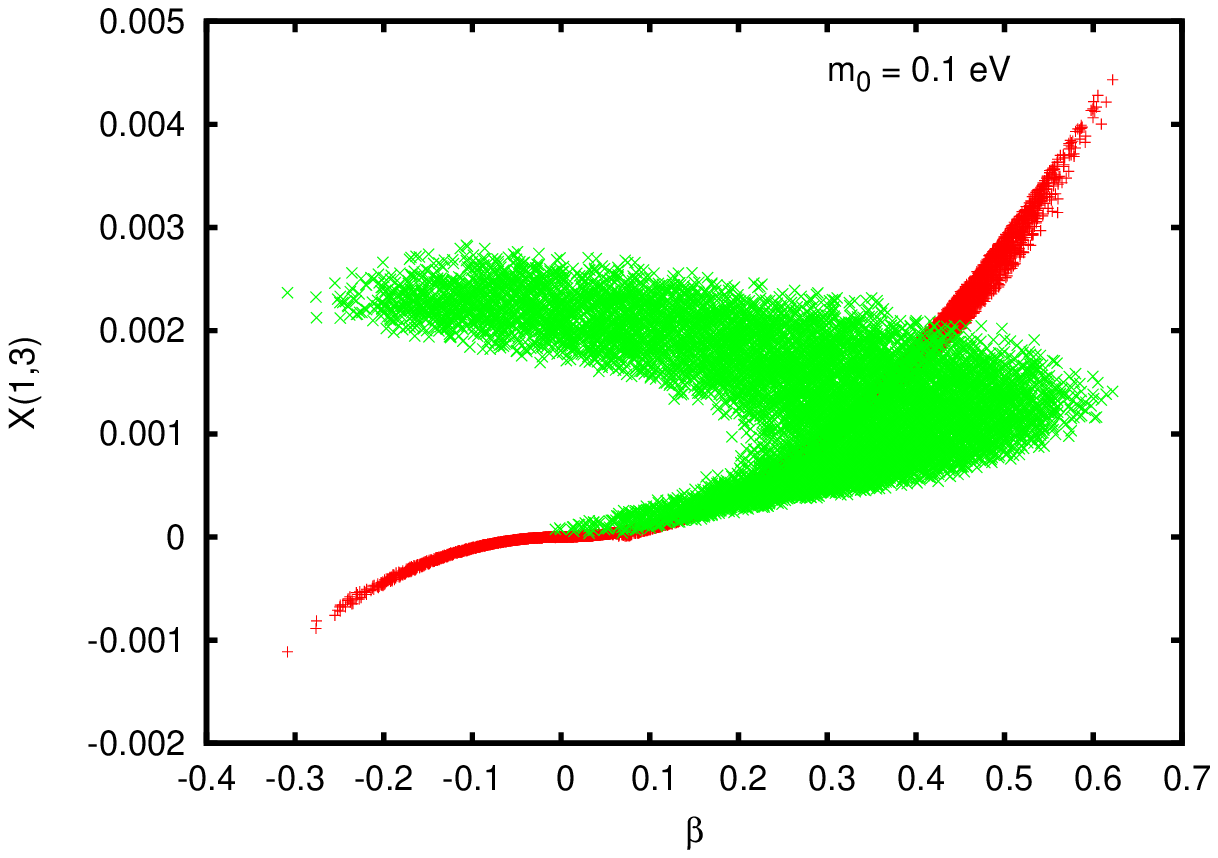}
\end{center}
\caption{Range of $X(1,3)$ as a function of $\beta$ for $\alpha = 0$ at $m_0 = 0.0,\,0.01,\,0.1\,{\rm eV}$, respectively.
The theory ranges are denoted by red whereas the experimental ranges from the bottom up approach are denoted by
green.}
\label{Xm0}
\end{figure}

Now we discuss the role of the non-zero lightest neutrino mass in determining the form of the perturbed matrix $V$.
In Table.~\ref{X1} and Table.~\ref{X2}, we report all the elements of the matrix $X$ for $m_0 = 0.01\,{\rm and}\,0.1\,{\rm eV}$ 
obtained using the bottom up approach for the TBM mixing scenario.
\begin{table}[ht]
\centering                          
\begin{tabular}{|c|c|c|c|c|c|c|}           
\hline\hline                        
TBM & X(1,1)/{\rm eV} & X(1,2)/{\rm eV} & X(1,3)/{\rm eV} & X(2,2)/{\rm eV} & X(2,3)/{\rm eV} & X(3,3)/{\rm eV} \\ [0.5ex]   
\hline                              
Central values & $1.02$ & $0.044$ & $0.26$ & $1.45$ & $0.68$ & $5.00$ \\               
$3\sigma$ range & $[1.0, 1.22]$ & $[-0.054, 0.086]$ & $[0.0079, 0.95]$ & $[1.31, 1.60]$ & $[-0.15, 0.96]$ &                                                                                         
$[4.68, 5.26]$ \\ [1ex]           
\hline
\end{tabular}
\caption{Values of $X (i,j)\times 10^{2}$ for TBM mixing from the bottom up approach for $m_0 = 0.01\,{\rm eV}$.}  
\label{X1}          
\end{table}
\begin{table}[ht]
\centering                          
\begin{tabular}{|c|c|c|c|c|c|c|}           
\hline\hline                        
TBM & X(1,1)/{\rm eV} & X(1,2)/{\rm eV} & X(1,3)/{\rm eV} & X(2,2)/{\rm eV} & X(2,3)/{\rm eV} & X(3,3)/{\rm eV} \\ [0.5ex]   
\hline                              
Central values & $10.0$ & $0.013$ & $0.075$ & $10.08$ & $0.21$ & $11.16$ \\               
$3\sigma$ range & $[10.0, 10.06]$ & $[-0.0012, 0.023]$ & $[0.00028, 0.28]$ & $[10.03, 10.11]$ & $[-0.046, 0.30]$ & 
$[11.02, 11.26]$ \\ [1ex]         
\hline
\end{tabular}
\caption{Values of $X (i,j)\times 10^{2}$ for TBM mixing from the bottom up approach for $m_0 = 0.1\,{\rm eV}$.}  
\label{X2}          
\end{table}
To demonstrate the dependency of $X(i,j)$ on $m_0$ we consider only one out of six set of solutions discussed above. In particular, we choose 
the most non-trivial one with $\alpha=0$ and $\beta,\, \gamma\neq 0$. Because from Fig.~\ref{Xbg} we see that, in this case, we can get the required 
values of $\theta_{13}$ even in the absence of perturbation in $(1,3)$ plane.  We fix the range of $\beta$, $\gamma$ from the experimental 
values of $X(1,2)$ and $X(2,2)$ given in Table.~\ref{X1} and Table.~\ref{X2} for two different values of $m_0$. We then check the consistency 
of remaining four elements. It is evident from Fig.~\ref{Xbg} that, in the limit $m_0=m_1\to 0$, the theoretical values of $X(1,3)$ has 
minimal overlapping with experimental values. Therefore, we choose this case to further study the dependency of $X(1,3)$ on non-zero values of 
$m_0$ and the result is shown in Fig.~\ref{Xm0}. It is clear that the fitting with the experimental data improves once we go from $m_0 = 0$ to non zero 
values of $m_0$. 

We summarize our findings in Table.~\ref{Y1}. It is clear from our analysis that any perturbation to the TBM mass matrix where two parameters are 
non zero can reproduce the experimental data. It is not necessary to have perturbation in $(1,3)$ plane to reproduce the experimental data as frequently 
discussed in the literature. Perturbations in $(1,2)$ and $(2,3)$ plane are equally good to get a reasonable fit within $3\sigma$ of the experimental data. 
However, once we set two parameters in $(1,2)$ and $(2,3)$ plane to zero, then perturbation along the $(1,3)$ plane is necessary to get compatible results 
with the experimental data. 

\begin{table}[htdp]
\begin{center}
\begin{tabular}{|c|c|c|}
\hline\hline
$\alpha = 0$ & $\beta,\, \gamma \ne 0$& $\surd$  \\[0.2cm]
\hline
$\beta = 0$ &$\alpha,\, \gamma \ne 0$ & $\surd$ \\[0.2cm]
\hline
$\gamma = 0$ &$\alpha,\, \beta \ne 0$  & $\surd$ \\[0.2cm]
\hline
$\alpha,\,\beta = 0$ &$\gamma \ne 0$  & $\times$ \\[0.2cm]
\hline
$\alpha,\,\gamma = 0$ &$\beta \ne 0$  & $\times$ \\[0.2cm]
\hline
$\gamma,\,\beta = 0$ &$\alpha \ne 0$  & $\surd$ \\[0.2cm]
\hline
\end{tabular}
\end{center}
\caption{Six different possibilities out of which four are consistent with data. These are true irrespective of the value of
the lightest neutrino mass $m_0$.}
\label{Y1}
\end{table}

\section{Conclusion}
\label{con}
The large value of $\theta_{13}$ predicted by Daya Bay and RENO put a stringent constraint on theoretical model buildings 
for neutrino mixing. All the well known mixing ansatzs such as TBM mixing, BM mixing, and DC mixing predict $\theta_{13}$ 
to be zero and hence not consistent with experiment. However, a lot of phenomenological models in the top-down scenario 
have been proposed in order to explain the non zero $\theta_{13}$, where a large value of the 1-3 mixing angle 
$\theta_{13}$ is generated through a perturbation to these mixing scenarios. In this paper we propose a perturbative 
bottom-up approach to quantify accurately the perturbation required for each element around a tree level mass matrix that 
is determined using the TBM mixing ansatz. Using the inputs from bottom-up approach we propose a model for neutrino 
masses and mixings. Though we used TBM mixing as an example, our analysis is more general and can be applied to any other 
ansatz which predicts $\theta_{13}=0$ at the tree-level.    

We first summarize our results for the bottom-up approach:
\begin{itemize}
\item
It is known that in a flavor basis the elements of the neutrino mass matrix can be written in terms of the oscillation parameters and the 
absolute value of lightest neutrino mass~($m_0$). In this context we point out that for $m_0 \ll 10^{-3}\,{\rm eV}$, 
the perturbed elements don't depend on the value of $m_0$. Therefore, all the perturbed elements can be determined exactly in terms 
of the oscillation parameters. However, in the opposite limit, where $m_0\gsim 10^{-3}\,{\rm eV}$, we find that the perturbed elements 
have exponential dependency on the value of $m_0$. We factor out the $m_0$ dependency of all $\delta M_\nu(i,j)$ using an exponential 
parameterization as given in Eq. (\ref{m_0}). This helps us in determining the exact dependency of the perturbed matrix elements on the 
value of $\theta_{13}$, $\theta_{23}$ and $\theta_{12}$ for all values of $m_0$. 
\smallskip
\item
In order to gauge the size of the perturbation to each element of the mass matrix, we first compare the perturbed matrix elements with 
the corresponding tree level mass matrix derived from the TBM mixing ansatz and find that only the $(1,2)$ and $(1,3)$ elements needs to 
be significantly modified to be consistent with the experimental data. This is true for normal as well as inverted ordering of the neutrino mass spectrum. 
\smallskip
\item
We show the exact dependency of the perturbed matrix elements on the value of $\theta_{13}$, $\theta_{23}$ and 
$\theta_{12}$. We find that $(1,1)$, $(1,2)$ and $(1,3)$ elements of the perturbed matrix do depend on all the three mixing angles. 
However, their dependency on $\theta_{12}$ is quite small as compared to that of $\theta_{13}$ and $\theta_{23}$. This, in act, is 
true for normal as well as inverted hierarchy of the neutrino mass spectrum. However, the $(2,2)$, $(2,3)$, and $(3,3)$ elements depend 
only on the value of $\theta_{23}$. 
Similar conclusions can be drawn for inverted hierarchy as well.
\end{itemize}

We use the results of bottom-up approach as guide-lines for determining a perturbative model of neutrino masses and mixings. We 
introduce a typical mixing matrix $U_M = U_{\rm TBM}\,V$ for the neutrino mass matrix, where $V (\alpha,\beta,\gamma)$ is the perturbed 
mixing matrix. We fix the parameters $\alpha$, $\beta$, and $\gamma$ from the oscillation data obtained using the bottom up approach, where 
$\alpha$, $\beta$ and $\gamma$ are the rotation angles in (1,3), (1,2) and (2,3) planes respectively. Here are some important observations 
from our top down analysis.
\begin{itemize}
\item
We find that we can set any one parameter to zero, i.e, $(\alpha,\,\beta,\,{\rm or}\,\gamma = 0)$ and still get a reasonably good fit with the 
experimental data that are derived using the bottom up approach. Although, keeping $\beta = 0$ and $\alpha,\,\gamma \ne 0$ gives a much better fit, 
it is worth mentioning that $\alpha = 0$ and $\beta,\,\gamma \ne 0$ also gives a reasonably good fit with the experimental data set. Thus, it is not 
necessary to have a perturbation in the $(1,3)$ plane in order to be compatible with the experimental data. This is one feature that was not 
explored in any earlier work.
\smallskip
\item
Once we set $(\alpha,\,\beta = 0)$ or $(\alpha,\,\gamma = 0)$, our results are not compatible within $3\sigma$ range of the experimental data. We, however, 
can set $\alpha \ne 0$ and $(\beta,\,\gamma = 0)$ and still get a reasonably good fit with the experimental data. This case has been explored extensively 
in the literature.
\smallskip
\item
It is worth mentioning that the lightest neutrino mass plays an important role in the parameter fitting. The fitting improves once we go from a 
zero to non zero values of lightest neutrino mass. In particular for $m_0 \gsim 10^{-3} {\rm eV}$ we get large overlapings between the theoretical 
values of $X(i,j)$ with their experimental values. 
\smallskip
\item
Since $\alpha$, $\beta$, and $\gamma$ can be set to zero separately, their ordering is not important. We further 
note that, if $V$ is parameterized by more than one variable then the perturbations are arbitrary. In other words, those can not 
be treated as true perturbations with respect to their tree level mixing angles.   
\end{itemize}
\vskip 1.0cm
\acknowledgements
RD and AKG would like to thank BRNS, Govt. of India for financial support.


\end{document}